\documentclass[12pt]{article}

\usepackage{times}
\usepackage{amsfonts}

  \topmargin - 0.5 cm 
\begin{document}

\baselineskip = 0.60 true cm

\noindent {\large \bf An angular momentum approach to 
quadratic Fourier transform, \\
Hadamard matrices, 
Gauss sums, 
mutually unbiased bases, \\
unitary group and Pauli group\footnote{Dedicated to the memory of Yurii Fedorovich Smirnov.}}\\

\vspace{0.5cm}

\noindent {\bf Maurice R Kibler}


\noindent {Universit\'e de Lyon, F--69622, Lyon, France \\
Universit\'e Lyon 1, Villeurbanne, France \\
and \\
CNRS/IN2P3, Institut de Physique Nucl\'eaire de Lyon, France} \\ \\

\noindent E-mail: m.kibler@ipnl.in2p3.fr

\vspace{1.25cm}

\noindent {\bf Abstract} 

\noindent The construction of unitary operator bases in a finite-dimensional Hilbert space 
is reviewed through a nonstandard approach combinining angular momentum theory and 
representation theory of $SU(2)$. A single formula for the bases is obtained from a polar 
decomposition of $SU(2)$ and analysed in terms of cyclic groups, quadratic Fourier transforms, 
Hadamard matrices and generalized Gauss sums. Weyl pairs, generalized Pauli operators and 
their application to the unitary group and the Pauli group naturally arise in this approach.

\vspace{1cm}



   \newpage

\section{Introduction}

Angular momentum theory \cite{BLangular} and its group-theoretical formulation 
in terms of the Wigner-Racah algebra of SU(2) \cite{WignerSR, Racah, BLWRa} 
(see also \cite{Sharp} for an extension to a finite or compact group)
are of central importance in subatomic, atomic, 
molecular and condensed matter physics. The components of any angular momentum 
(spin, isopin, orbital angular momentum, etc.) generate the Lie algebra of the 
group $SU(2)$. Therefore, $SU(2)$ 
and its noncompact extension $SU(1,1)$ are basic ingredients 
for dealing with generalized angular momenta. Chains of groups ending with 
$SO(3) \simeq SU(2) / \mathbb{Z}_2(+)$ or $SO(3) \subset SO(2)$ are of interest 
in subatomic and atomic physics. In this direction, one can mention the group 
$SU(3) \otimes SU(2) \otimes U(1)$ (related to the chain 
$U(3) \subset SU(2) \otimes U(1) \subset U(1)$) and its grand unified and/or 
supersymmetric extensions for describing elementary particles and their (strong 
and electroweak) interactions \cite{afterSharp}. Furthermore, one know the relevance in 
atomic physics of the chain $U(7) \subset SO(7) \subset G_2 \subset SO(3) \subset SO(2)$ 
for the electronic spectroscopy of $f^N$ ions \cite{Racah}. On the other side, chains ending
with $SU(2) \subset G$, where $G$ is a finite group (or a chain involving finite groups), 
proved to be of considerable interest in molecular and condensed matter spectroscopy 
\cite{Russes, Chinois, Antigonish79}. Recently, chains of type $SU(2) \subset G$ were also 
used in attempts to understand the flavor structure of quarks and leptons \cite{EMA}. The 
groups $SU(2)$ and $SU(1,1)$, as well as their $q$- or $qp$-deformations in the sense of Hopf 
algebras (see for instance \cite{SRR, 3SmirKib}, thus play a pivotal role in many areas of 
physical sciences. 

The representation theory of $SU(2)$ is generally adressed in two different ways. The standard 
one amounts to diagonalise the complete set $\{ j^2 , j_z \}$ involving the Casimir operator $j^2$ 
and one generator $j_z$ of $SU(2)$. Another way is to consider a set $\{ j^2 , v \}$, where $v$ 
is an operator defined in the enveloping algebra of $SU(2)$ and invariant under a subgroup of 
$SU(2)$. A third way (not very well-known) consists in diagonalising a complete set 
$\{ j^2 , v_{ra} \}$, where $v_{ra}$ stands for a two-parameter operator which commutes with 
$j^2$ and is a pseudoinvariant under a cyclic group \cite{Kib99-01-05}. 

It is the aim of this review paper to show that the third approach to the representation 
theory of $SU(2)$ opens a window on the apparently disconnected subjects enumerated in the title. 

The plan of the paper is as follows. The minimal requirements for a $\{ j^2 , v_{ra} \}$ approach to $SU(2)$ 
(i.e., a nonstandard approach to angular momentum theory) are given in Section 2 and in two appendices. 
Section 3 deals with quadratic sums (in relation with quadratic discrete Fourier transforms, generalized 
Hadamard matrices, generalized quadratic Gauss sums and mutually unbiased bases) and Section 4 is devoted 
to unitary groups and Pauli groups.  

The present paper is dedicated to the memory of the late Professor Yurii Fedorovich Smirnov who contributed 
to many domains of mathematical physics (e.g., Lie groups and Lie algebras, quantum groups, special functions) 
and theoretical physics (e.g., nuclear, atomic and molecular physics, crystal- and ligand-field theory).  

A few words about some of the notations is in order. The bar indicates complex conjugation. The 
symbol $\delta_{a , b}$ stands for the Kronecker symbol of $a$ and $b$. We use $I$ and $I_d$ to 
denote the identity operator and the $d$-dimensional unity matrix, respectively. The operator 
$A^{\dagger}$ stands for the adjoint of the operator $A$. We note as $[A , B]_-$ and $[A , B]_+$ 
the commutator and the anticommutator of the operators $A$ and $B$, respectively. We use the 
Dirac notation $| \psi \rangle$ for a vector in an Hilbert space; furthermore, 
$\langle \phi | \psi \rangle$ and $| \phi \rangle \langle \psi |$ are respectively the inner 
product and the outer product of the vectors $| \psi \rangle$ and $| \phi \rangle$. 
The symbols 
$\oplus$ and $\ominus$ stand respectively for the addition and subtraction modulo $d$ while 
$\otimes$ and $\uplus$ are used respectively for the direct product of vectors or operators and
the direct sum of vector spaces. The matrices of type $E_{\lambda,\mu}$ with the matrix elements 
    \begin{eqnarray}	                  
\left( E_{\lambda,\mu} \right)_{\lambda',\mu'} := \delta_{\lambda , \lambda'} \delta_{\mu , \mu'}
    \label{definition de E} 
	  \end{eqnarray} 
stand for generators of the Lie group $GL(d , \mathbb{C})$. For $a$ and $b$ coprime, we take 
          \begin{eqnarray}
\pmatrix{
a \cr
b \cr
}_L := \cases{
               +1  \ {\rm if}  \ a     = k^2  \ {\rm mod}(b)       
\cr \cr
               -1  \ {\rm si}  \ a \not= k^2  \ {\rm mod}(b)      
                  }
          \end{eqnarray}
to denote the Legendre symbol of $a$ and $b$ (equal to 1 
if $a$ is     a quadratic residu modulo $b$ and $-1$ 
if $a$ is not a quadratic residu modulo $b$). In addition, the 
integer inverse $(a \backslash b)$ of $a$ with respect to $b$ is given by 
          \begin{eqnarray}
a (a \backslash b) = 1 \ {\rm mod}(b). 
          \end{eqnarray}
Finally, the $q$-deformed number $\left[ n \right]_q$ and the $q$-deformed factorial $\left[ n \right]_q !$, 
with $n \in \mathbb{N}$, are defined by 
          \begin{eqnarray}
  \left[ n \right]_q := \frac{1-q^n}{1-q}
          \end{eqnarray}
and
          \begin{eqnarray}
  \left[ n \right]_q! := 
  \left[ 1 \right]_q 
  \left[ 2 \right]_q  \ldots 
  \left[ n \right]_q \qquad \left[ 0 \right]_q! := 1
          \end{eqnarray}
where $q$ is taken is this paper as a primitive root of unity.

\section{A nonstandard approach to $su(2)$}

In some previous works \cite{Kib99-01-05}, we developed a 
nonstandard approach to the Lie algebra $su(2)$ and studied the corresponding Wigner-Racah 
algebra of the group $SU(2)$. This nonstandard approach is based on a polar 
decomposition of $su(2)$, based in turn on a troncated oscillator algebra (see Appendices A and B). It 
yields nonstandard bases for the irreducible representations of $SU(2)$ and new Clebsch-Gordan 
coefficients for the angular momentum theory. Basically, the 
approach amounts to replace the set $\{ j^2 , j_z \}$, familiar in 
quantum mechanics, by a set $\{ j^2 , v_{ra} \}$ ($j^2$ and $j_z$ are the Casimir 
operator and the Cartan generator of $su(2)$, respectively). 

The operator $v_{ra}$ acts on the $(2j+1)$-dimensional subspace ${\cal E}(2j+1)$, 
associated with the angular momentum $j$, of the representation space of $SU(2)$. We define it here by 
          \begin{eqnarray}
          v_{ra} := {e}^{{i} 2 \pi j r} |j , -j \rangle \langle j , j| 
                  + \sum_{m = -j}^{j-1} q^{(j-m)a} |j , m+1 \rangle \langle j , m| 
          \label{definition of vra} 
          \end{eqnarray}
where  
          \begin{eqnarray}
          q := \exp \left( {2 \pi {i} \over 2j+1} \right) \qquad 
          2j \in \mathbb{N} \qquad 
          r \in \mathbb{R} \qquad 
          a \in \mathbb{Z}_{2j+1}
          \label{parameters} 
          \end{eqnarray}
and, for fixed $j$, the vectors $|j , m \rangle$ (with $m = j, j-1, \ldots, -j$) satisfy the 
eigenvalue equations
          \begin{eqnarray}
          j^2 |j , m \rangle = j(j+1) |j , m \rangle \qquad 
          j_z |j , m \rangle = m      |j , m \rangle
          \end{eqnarray}
familiar in angular momentum theory. The vectors $|j , m \rangle$ span the Hilbert space 
${\cal E}(2j+1) \sim \mathbb{C}^{2j+1}$ and are taken in an orthonormalized form with 
          \begin{eqnarray}
          \langle j, m |j , m' \rangle = \delta_{m,m'}.
          \end{eqnarray}
Obviously, the operator $v_{ra}$ is unitary and commutes with $j^2$. The 
spectrum of the set $\{ j^2 , v_{ra} \}$ is described by 

\bigskip

\noindent {\bf Result 1.} {\it For fixed $j$, $r$ and $a$, the $2j+1$ vectors 
          \begin{eqnarray}
|j \alpha ; r a \rangle := \frac{1}{\sqrt{2j+1}} \sum_{m = -j}^{j} 
q^{(j + m)(j - m + 1)a / 2 - j m r + (j + m)\alpha} | j , m \rangle
          \label{j alpha r a in terms of jm}
          \end{eqnarray} 
with $\alpha = 0, 1, \ldots, 2j$, are common eigenvectors of $v_{ra}$ and $j^2$. The 
eigenvalues of $v_{ra}$ and $j^2$ are given by 
      \begin{eqnarray}
v_{ra} |j \alpha ; r a \rangle = q^{j(a+r) - \alpha} |j \alpha ; r a \rangle \quad 
j^2    |j \alpha ; r a \rangle = j(j+1)              |j \alpha ; r a \rangle \quad
\alpha = 0, 1, \ldots, 2j. 
      \label{evp de vra}
      \end{eqnarray}
The spectrum of $v_{ra}$ is nondegenerate.}

\bigskip

The set $\{ |j \alpha ; r a \rangle : \alpha = 0, 1, \ldots, 2j \}$ constitutes another 
orthonormal basis, besides the basis $\{ |j , m \rangle : m = j, j-1, \ldots, -j \}$, 
of ${\cal E}(2j+1)$ in view of 
      \begin{eqnarray}
\langle j \alpha ; r a | j \beta ; r a \rangle = \delta_{\alpha,\beta}.
      \label{jalphabetara}
      \end{eqnarray}
Note that the value of $\langle j \alpha ; r a | j \beta ; s b \rangle$ is much more 
involved for $r \not= s$ and $a \not= b$ and needs the calculation of Gauss sums 
as we shall see below.

The Wigner-Racah algebra of $SU(2)$ can be developed in the $\{ j^2, v_{ra} \}$ scheme. 
This leads to Clebsch-Gordan coefficients and (3~--~$j \alpha$)$_{ra}$ symbols with 
properties very different from the ones of the usual $SU(2) \subset U(1)$ Clebsch-Gordan 
coefficients and 3~--~$jm$ symbols corresponding to the $\{ j^2, j_z \}$ scheme \cite{Kib99-01-05}.       
          
The nonstandard approach to angular momentum theory briefly summarized above is 
especially useful in quantum chemistry for problems involving cyclic symmetry. This 
is the case for a ring-shape molecule with $2j+1$ atoms at the vertices of a regular 
polygon with $2j+1$ sides or for a one-dimensional chain of $2j+1$ spins ($\frac{1}{2}$-spin 
each) \cite{AlbouyKib}. In this connection, we observe that the vectors of type $|j \alpha ; r a \rangle$ 
are specific symmetry-adapted vectors \cite{Melvin, 2Altmann}. Symmetry-adapted vectors
are widely used in quantum 
chemistry, molecular physics and condensed matter physics as for instance in rotational 
spectroscopy of molecules \cite{Dijon} and ligand-field theory \cite{LFT}. However, 
the vectors $| j \alpha ; r a \rangle$ differ from the 
symmetry-adapted vectors considered in Refs.~\cite{Moret, JMSCRAS, PatWin, Michel77} in the sense 
that $v_{ra}$ is not an invariant under some finite subgroup (of crystallographic interest) 
of the orthogonal group $O(3)$. Indeed, $v_{ra}$ is a pseudoinvariant \cite{ZhiKib} under
the Wigner operator $P_{R(\varphi)}$ associated with the rotation $R(\varphi)$, 
around the quantization axis $Oz$, with the angle
      \begin{eqnarray}
\varphi := p \frac{2 \pi}{2j+1} \qquad p = 0, 1, \ldots, 2j
      \end{eqnarray}
since 
      \begin{eqnarray}
P_{R(\varphi)} v_{ra} P_{R(\varphi)}^{\dagger} = e^{-i \varphi} v_{ra}.
      \label{trans}
      \end{eqnarray}         
More precisely, we have 

\bigskip

\noindent {\bf Result 2.} {\it The operator 
$v_{ra}$ transforms according to an irreducible representation of the cyclic subgroup 
$C_{2j+1} \sim \mathbb{Z}_{2j+1}(+)$ of the special orthogonal group $SO(3)$. In terms 
of vectors, one has 
      \begin{eqnarray}
P_{R(\varphi)} |j \alpha ; r a \rangle = q^{jp} |j \beta ; r a \rangle \qquad \beta := \alpha \ominus p
      \end{eqnarray}            
so that the set $\{ |j \alpha ; r a \rangle : \alpha = 0, 1, \ldots, 2j \}$ is stable under $P_{R(\varphi)}$. 
The latter set spans the regular representation of $C_{2j+1}$.}

\section{Variations on quadratic sums}

     \subsection{Quadratic discrete Fourier transform}

We leave the domain of angular momentum theory and adopt the following notations 
      \begin{eqnarray}
d := 2j + 1 \qquad k := j - m \qquad | k \rangle := | j , m \rangle.
      \label{notations}
      \end{eqnarray}  
These notations are particularly adapted to quantum information and quantum 
computation. In these new notations, we have 
          \begin{eqnarray}
          v_{ra} = {e}^{{i} \pi (d-1) r} |d-1 \rangle \langle 0| 
                  + \sum_{k=1}^{d-1} q^{ka} |k-1 \rangle \langle k|.
          \label{definition of vra en notations qiqc} 
          \end{eqnarray}
From now on, we assume that $d \geq 2$ and $r=0$ (the case $d = 1$ and $r \not= 0$, 
although of interest in the theory of angular momentum, is not essential for what follows).  
In addition, we put 
          \begin{eqnarray}
| a \alpha \rangle := | j \alpha ; 0 a \rangle
          \label{definition de a alpha} 
          \end{eqnarray}
with $a$ and $\alpha$ in the ring $\mathbb{Z}_d := \mathbb{Z}/d\mathbb{Z}$. Then, 
Eq.~(\ref{evp de vra}) gives
      \begin{eqnarray}
v_{0a} |a \alpha \rangle = q^{(d-1)a/2 - \alpha} |a \alpha \rangle
      \label{evp de vra en notation qiqc}
      \end{eqnarray}
with
    \begin{eqnarray}	                  
|a \alpha \rangle = \frac{1}{\sqrt{d}} \sum_{k = 0}^{d-1} 
q^{(k + 1)(d - k - 1)a / 2 - (k + 1) \alpha} | k \rangle. 
    \label{vecteur a alpha} 
	  \end{eqnarray} 
Equation (\ref{vecteur a alpha}) can be rewritten as 
    \begin{eqnarray}	                  
|a \alpha \rangle = \sum_{k = 0}^{d-1} \left( F_a \right)_{k \alpha} | k \rangle  
    \label{vecteur a alpha en Fa} 
	  \end{eqnarray} 
where
    \begin{eqnarray}	                  
\left( F_a \right)_{k \alpha} := \frac{1}{\sqrt{d}} 
q^{(k + 1)(d - k - 1)a / 2 - (k + 1) \alpha} 
    \label{matrice Fa} 
	  \end{eqnarray} 
is the $k \alpha$-th matrix element of a $d \times d$ matrix $F_a$ (the 
matrix $F_a$ can be seen as the matrix associated with the character 
table of the cyclic group $C_d$ pre- and post-multiplied by diagonal matrices).

Equations (\ref{vecteur a alpha en Fa})-(\ref{matrice Fa}) define a quadratic quantum Fourier transform. The matrix $F_a$ is unitary so that (\ref{vecteur a alpha en Fa}) 
can be inverted to give
    \begin{eqnarray}	                  
| k \rangle = \sum_{\alpha = 0}^{d-1} \overline{\left( F_a \right)_{k \alpha}} | a \alpha \rangle 
    \label{vecteur k en Fa} 
	  \end{eqnarray}
or
    \begin{eqnarray}	                  
| k \rangle = \frac{1}{\sqrt{d}} \sum_{\alpha = 0}^{d-1} q^{-(k + 1)(d - k - 1)a / 2 + (k + 1) \alpha} | a \alpha \rangle.  
    \label{vecteur k} 
	  \end{eqnarray}
In the special case $a = 0$, we have 
    \begin{eqnarray}	                  
|0 \alpha \rangle = q^{ - \alpha} \frac{1}{\sqrt{d}} \sum_{k = 0}^{d-1} 
e^{ - \frac{2 \pi i}{d} \alpha k } | k \rangle \Leftrightarrow 
| k \rangle = \frac{1}{\sqrt{d}} \sum_{\alpha = 0}^{d-1} 
e^{ \frac{2 \pi i}{d} (k+1) \alpha } | 0 \alpha \rangle. 
	  \end{eqnarray} 
Consequently, the quadratic quantum Fourier transform reduces to the ordinary quantum 
Fourier transform (up to a phase factor). The corresponding matrix $F_0$ satisfies
    \begin{eqnarray}	                  
F_0^4 = q I_d
    \label{F0 puissane 4} 
	  \end{eqnarray} 
to be compared to the well-known relation $F^4 = I_d$ for the standard quantum Fourier 
transform \cite{Vourdas04}. 

At this stage, we forsee that $d+1$ (orthonormal) bases of the space ${\cal E}(d)$ play an 
important role in the present paper: (i) the basis 
     \begin{eqnarray}
B_d := \{ |j , m \rangle : m = j, j-1, \ldots, -j  \} \Leftrightarrow 
B_d := \{ | k \rangle    : k = 0, 1,   \ldots, d-1 \} 
     \label{canonical basis}
     \end{eqnarray}
associated with the $\{ j^2, j_z \}$ scheme, 
known as the spherical or canonical basis in the theory of angular momentum, 
and as the computational basis in quantum information and quantum computation and 
(ii) the $d$ bases  
     \begin{eqnarray}
B_a := \{ | a \alpha \rangle : \alpha = 0, 1, \ldots, d-1 \} \qquad a = 0, 1, \ldots, d-1    
     \label{definition B_a}
     \end{eqnarray}
(noted $B_{0a}$ in Ref.~\cite{KibJPhysA}) associated with the $\{ j^2, v_{0a} \}$ scheme. 

To close this subsection, let us show how the preceding developments can be used for 
defining a quadratic discrete Fourier transform. We start from the formal transformation
    \begin{eqnarray}	                  
x := \{ x(k     ) \in \mathbb{C} : k      = 0, 1, \ldots, d-1 \} \to 
y := \{ y(\alpha) \in \mathbb{C} : \alpha = 0, 1, \ldots, d-1 \} 
	  \end{eqnarray} 
defined via
    \begin{eqnarray}	                  
y(\alpha) := \frac{1}{\sqrt{d}} \sum_{k = 0}^{d-1} 
q^{(k + 1)(d - k - 1)a / 2 - (k + 1) \alpha} x(k) 
    \label{transfo xy} 
	  \end{eqnarray} 
where $a$ can take any of the values $0, 1, \ldots, d-1$. Alternatively, for fixed $a$ 
we have 
    \begin{eqnarray}	                  
y(\alpha) = \sum_{k = 0}^{d-1} \left( F_a \right)_{k \alpha} x(k) \qquad \alpha = 0, 1, \ldots, d-1.
    \label{transfo xy en Fa} 
	  \end{eqnarray} 
The inverse transformation $y \to x$ is described by 
    \begin{eqnarray}	                  
x(k) = \sum_{\alpha = 0}^{d-1} \overline{\left( F_a \right)_{k \alpha}} y(\alpha) \qquad k = 0, 1, \ldots, d-1.
    \label{transfo yx en Fa} 
	  \end{eqnarray} 
The bijective transformation $x \leftrightarrow y$ can be thought of as a quadratic 
discrete Fourier transform. The case  $a = 0$  corresponds to the ordinary discrete 
Fourier transform (up to a phase factor). These matters lead to the following result 
which generalizes the Parseval-Plancherel theorem for the ordinary discrete Fourier 
transform. 

\bigskip

\noindent {\bf Result 3.} {\it The quadratic discrete Fourier transforms 
$x \leftrightarrow y$ and $x' \leftrightarrow y'$ associated withe same matrix 
matrix $F_a$, $a \in \mathbb{Z}_d$, satisfy the conservation rule 
    \begin{eqnarray}	                  
\sum_{\alpha = 0}^{d-1} \overline{y(\alpha)} y'(\alpha) = \sum_{k = 0}^{d-1} \overline{x(k)} x'(k)
    \label{Parseval-Plancherel} 
	  \end{eqnarray}
where the common value is independent of $a$.}

     \subsection{Generalized Hadamard matrices}

The modulus of each matrix element of $F_a$ (with $a \in \mathbb{Z}_d$) is equal 
to $1/\sqrt{d}$. Therefore, the unitary matrix $F_a$ turns out to be a generalized 
Hadamard matrix. We adopt here the following definition. A $d \times d$ generalized 
Hadamard matrix is a unitary matrix whose each entry has a modulus equal to 
$1/\sqrt{d}$ \cite{Dita}. Note that the latter normalization, used in quantum 
information \cite{Bengtsson-et-al, Weigert}, differs 
from the usual one according to which a $d \times d$ generalized Hadamard matrix 
$H$ is a complex matrix such that $H^{\dagger} H = d I_d$ and for which the modulus 
of each element is 1 \cite{TadejandZ}. In this respect, the generalized Hadamard matrix 
$H_a$ considered in \cite{AlbouyKib} corresponds to $\sqrt{d} F_a$ up to permutations.

\bigskip

\noindent {\bf Example 1.} By way of illustration, from (\ref{matrice Fa}) we get the familiar Hadamard matrices 
     \begin{eqnarray}
F_0 = \frac{1}{\sqrt{2}} 
\pmatrix{
  1     &-1   \cr
  1     &1    \cr
} \qquad 
F_1 = \frac{1}{\sqrt{2}} 
\pmatrix{
  i     &-i  \cr
  1     &1   \cr
}
     \end{eqnarray}
for $d=2$ and 
      \begin{eqnarray}
F_0 = \frac{1}{\sqrt{3}} 
\pmatrix{
  1     &\omega^2  &\omega   \cr
  1     &\omega    &\omega^2 \cr
  1     &1         &1        \cr  
} \
F_1 = \frac{1}{\sqrt{3}} 
\pmatrix{
\omega  &1        &\omega^2   \cr
\omega  &\omega^2  &1          \cr
1       &1         &1          \cr  
} \
F_2 = \frac{1}{\sqrt{3}} 
\pmatrix{
\omega^2  &\omega       &1              \cr
\omega^2  &1            &\omega         \cr
1         &1            &1              \cr  
}
     \end{eqnarray}
(with $\omega := e^{i 2 \pi / 3}$) for $d = 3$. Another example is
      \begin{eqnarray}
F_0 = \frac{1}{\sqrt{6}} 
\pmatrix{
  1     &\tau     &\tau^2  &-1    &-\tau    &-\tau^2 \cr
  1     &\tau^2   &-\tau   &1     &\tau^2   &-\tau   \cr
  1     &-1       &1       &-1    &1        &-1      \cr
  1     &-\tau    &\tau^2  &1     &-\tau    &\tau^2  \cr
  1     &-\tau^2  &-\tau   &-1    &\tau^2   &\tau    \cr
  1     &1        &1       &1     &1        &1       \cr
}
     \end{eqnarray}
(with $\tau := e^{- i \pi / 3}$) which readily follows from (\ref{matrice Fa}) for 
$d = 6$ and $a=0$.

\bigskip

We sum up and complete this section with the following result (see also \cite{AlbouyKib, KibIJMPB}).

\bigskip

\noindent {\bf Result 4.} {\it The matrix 
    \begin{eqnarray}	                  
F_a = \frac{1}{\sqrt{d}} \sum_{k=0}^{d-1} \sum_{\alpha=0}^{d-1}
q^{(k + 1)(d - k - 1)a / 2 - (k + 1) \alpha} E_{k,\alpha}
    \label{Hadamard} 
	  \end{eqnarray} 
associated with the quadratic quantum Fourier transform (\ref{vecteur a alpha en Fa})
is a $d \times d$ generalized Hadamard matrix. It reduces the endomorphism 
associated with the operator $v_{0a}$: 
      \begin{eqnarray}
F_a^{\dagger} V_{0a} F_a = q^{(d-1)a / 2} \sum_{\alpha = 0}^{d-1} q^{-\alpha} E_{\alpha , \alpha} = 
q^{(d-1)a / 2} \pmatrix{
1                    &      0 &       \ldots &          0    \cr
0                    & q^{-1} &       \ldots &          0    \cr
\vdots               & \vdots &       \ldots &     \vdots    \cr
0                    &      0 &       \ldots & q^{-(d-1)}    \cr
}
     \label{endomor}
     \end{eqnarray}
where the matrix
      \begin{eqnarray}
V_{0a} := \sum_{k = 0}^{d-1} q^{ka} E_{k \ominus 1 , k} =
\pmatrix{
0                    &    q^a &      0  & \ldots &          0    \cr
0                    &      0 & q^{2a}  & \ldots &          0    \cr
\vdots               & \vdots & \vdots  & \ldots &     \vdots    \cr
0                    &      0 &      0  & \ldots & q^{(d-1)a}    \cr
1                    &      0 &      0  & \ldots &          0    \cr
}
     \label{matrix V0a}
     \end{eqnarray}
represents the linear operator $v_{0a}$ on the basis $B_d$.} 

     \subsection{Generalized quadratic Gauss sums}
     
The Hadamard matrices $F_a$ and $F_b$ ($a , b \in \mathbb{Z}_d$) 
are connected to the inner product $\langle a \alpha | b \beta \rangle$. In fact, we have 
     \begin{eqnarray}
\langle a \alpha | b \beta \rangle = \left( F_a^{\dagger} F_b \right)_{\alpha \beta}.
     \label{inner product en F}
     \end{eqnarray}
A direct calculation yields 
     \begin{eqnarray}
\langle a \alpha | b \beta \rangle = 
\frac{1}{d} \sum_{k = 0}^{d-1} q^{k(d-k)(b-a) / 2 - k(\beta - \alpha)} 
     \label{bb}
     \end{eqnarray}
or
     \begin{eqnarray}
\langle a \alpha | b \beta \rangle = \frac{1}{d} \sum_{k = 0}^{d-1} 
e^{i \pi \{ (a-b)k^2 + [d(b-a) + 2(\alpha-\beta)]k \} / d}.
     \label{a alpha b beta en expo}
     \end{eqnarray}
Hence, each matrix element of $F_a^{\dagger} F_b$ can be put in the form of 
a generalized quadratic Gauss sum $S(u, v, w)$ defined by \cite{BerndtEW} 
     \begin{eqnarray}
     S(u, v, w) := \sum_{k = 0}^{|w|-1} e^{i \pi (u k^2 + v k) / w}
     \label{dd}
     \end{eqnarray}
where $u$, $v$ and $w$ are integers such that $u$ and $w$ are mutually prime, 
$uw \not= 0$ and $uw + v$ is even. In detail, we obtain
     \begin{eqnarray}
\langle a \alpha | b \beta \rangle = \left( F_a^{\dagger} F_b \right)_{\alpha \beta} = \frac{1}{d} S(u, v, w)   
     \label{ee}
     \end{eqnarray}
with the parameters
     \begin{eqnarray}
     u = a - b \qquad v = -(a - b)d + 2(\alpha - \beta) \qquad w = d
     \label{ff}
     \end{eqnarray}
which ensure that $uw + v$ is necessarily even.

In the particular case $d = 2$ (of special interest for qubits), we directly get 
     \begin{eqnarray}
\langle a \alpha | b \beta \rangle = \frac{1}{2} \left[ 1 + e^{i \pi (b - a + 2 \alpha - 2 \beta) / 2} \right]
     \label{cas d vaut 2}   
     \label{gg}
     \end{eqnarray}
which reduces to 
     \begin{eqnarray}
\langle a \alpha | a \beta \rangle = \delta_{\alpha , \beta} \qquad b=a,  
     \label{xx}
     \end{eqnarray}
and
     \begin{eqnarray}
\langle a \alpha | b \beta \rangle = \frac{1}{2} (1 \pm i) \qquad b \not= a,     
     \label{yy}
     \end{eqnarray}
where the $+$ sign corresponds to $b-a + 2(\alpha-\beta) =  1, -3$ 
  and the $-$ sign             to $b-a + 2(\alpha-\beta) = -1,  3$. 

In the general case $d$ arbitrary (of interest for qudits), the calculation of  
$S(u, v, w)$ can be achieved by using the methods described in \cite{BerndtEW} (see also 
\cite{Hannay, Matsutani, Rosu, Merkel}). The cases of interest for what follows are 
($u$ even, $v$ even, $w$  odd),
($u$  odd, $v$  odd, $w$  odd) and 
($u$  odd, $v$ even, $w$ even). This leads to 

\bigskip

\noindent {\bf Result 5.} {\it  
For $a \not=b$, $d$ arbitrary and $u+v+w$ odd, the inner product $\langle a \alpha | b \beta \rangle$ 
and the $\alpha \beta$-th element of the matrix $F_a^{\dagger} F_b$ follow from 

\underline{case $u = a-b$ even, $v = d(b-a) + 2(\alpha - \beta)$ even, $w=d$ odd}:
     \begin{eqnarray}
\langle a \alpha | b \beta \rangle = \left( F_a^{\dagger} F_b \right)_{\alpha \beta} = \sqrt{\frac{1}{w}} 
\pmatrix{
u \cr
w \cr
}_L
\exp { \left( - i \frac{\pi}{4} [w-1 + \frac{u}{w} (u \backslash w)^2 v^2] \right) }
     \label{cas eeo}
     \end{eqnarray}

\underline{case $u = a-b$ odd, $v = d(b-a) + 2(\alpha - \beta)$ odd, $w=d$ odd}:   
     \begin{eqnarray}
\langle a \alpha | b \beta \rangle = \left( F_a^{\dagger} F_b \right)_{\alpha \beta} = \sqrt{\frac{1}{w}} 
\pmatrix{
u \cr
w \cr
}_L
\exp { \left( - i \frac{\pi}{4} [w-1 + 16 \frac{u}{w} (4u \backslash w)^2 v^2] \right) }
     \label{cas ooo}
     \end{eqnarray}   

\underline{case $u = a-b$ odd, $v = d(b-a) + 2(\alpha - \beta)$ even, $w=d$ even}:        
     \begin{eqnarray}
\langle a \alpha | b \beta \rangle = \left( F_a^{\dagger} F_b \right)_{\alpha \beta} = \sqrt{\frac{1}{w}} 
\pmatrix{
w \cr
u \cr
}_L
\exp { \left( - i \frac{\pi}{4} u [-1 + \frac{1}{w} (u \backslash w)^2 v^2] \right) }     
     \label{cas oee}
     \end{eqnarray}   
so that the matrix $F_a^{\dagger} F_b$ is a Hadamard matrix for each case under consideration.}

\bigskip

Finally, for $a = b$ and $d$ arbitrary we recover the orthonormality property 
(see (\ref{jalphabetara}))
     \begin{eqnarray}
\langle a \alpha | a \beta \rangle = \delta_{\alpha , \beta}
     \label{aaa}
     \end{eqnarray}  
from a direct calculation of the right-hand side of (\ref{a alpha b beta en expo}). 
   
     \subsection{Mutually unbiased bases}

Speaking generally, two $d$-dimensional bases 
$B_a = \{ | a \alpha \rangle : \alpha \in \mathbb{Z}_d \}$ and 
$B_b = \{ | b \beta  \rangle : \beta  \in \mathbb{Z}_d \}$ 
are said to be mutually unbiased if and only if 
          \begin{eqnarray}
| \langle a \alpha | b \beta \rangle | = \delta_{a , b}
\delta_{\alpha , \beta} + (1 - \delta_{a , b}) \frac{1}{\sqrt{d}}
          \label{definition des mubs}
          \end{eqnarray}          
for any $\alpha$ and $\beta$ in the ring $\mathbb{Z}_d$. It is well-known that the number of 
mutually unbiased bases (MUBs) in the Hilbert space $\mathbb{C}^d$ cannot be greater than 
$d+1$ \cite{Delsarte, Ivanovic, WoottersFields, Kerdock}. In fact, the maximum number $d+1$ 
is attained when $d$ is the power of a prime number \cite{WoottersFields, Kerdock}. Despite 
a considerable amount of works, the maximum number of MUBs is unknown 
when $d$ is not a power of a prime. In this respect, several numerical studies 
strongly suggest that there are only three MUBs for $d=6$ 
(see for example \cite{Bengtsson-et-al, Weigert, Zauner, Grassl, Weigert2}). MUBs 
are closely connected with the concept of complementarity in quantum 
mechanics. There are of paramount importance in classical information 
theory (Kerdock codes and network communication protocols) \cite{Kerdock, network}, in 
quantum information theory (quantum cryptography and quantum state tomography) \cite{Cerf}
and in the solution of the Mean King problem 
\cite{Englert, Aravind, rab5-Roi, rab6-Roi,rab7-Roi,rab8-Roi}. Recently, 
it was pointed out and confirmed that MUBs are also of central importance 
in the formalism of Feynman path integrals \cite{Svetlichny, Tolar2009}. 
Finally, it should be emphasized that the concept of MUBs also exists 
in infinite dimension \cite{MUBscontinues}. There are numerous ways of constructing 
sets of MUBs. Most of them are based on discrete Fourier analysis over Galois fields and 
Galois rings, discrete Wigner functions, generalized Pauli matrices, mutually orthogonal 
Latin squares, finite geometry methods and Lie-like approaches (see Refs. 
\cite{AlbouyKib, KibJPhysA, Bengtsson-et-al, Weigert, KibIJMPB, Delsarte, Ivanovic, WoottersFields, Kerdock, Zauner, Grassl, Weigert2, network, Cerf, Englert, Aravind, rab5-Roi, rab6-Roi,rab7-Roi,rab8-Roi, BandyoGPM5, LawrenceGPM6, Chaturvedi, Vlasov, Klappenecker, Gibbons, Pittenger, Durt, Archer, Aschbacher, Sulc} for an nonexhaustive list of references). 

\subsubsection{Case $d$ prime} 
We have the following important result. (See also \cite{Sulc} for a 
recent alternative group-theoretical approach to the case $d$ prime.)

\bigskip

\noindent {\bf Result 6.} {\it In the case where $d=p$ is a prime number (even or odd), 
one has                          
     \begin{eqnarray}
| \langle a \alpha | b \beta \rangle | = 
\left| \left( F_a^{\dagger} F_b \right)_{\alpha \beta} \right| = \frac{1}{\sqrt{p}} \qquad a \not= b   
     \label{cas d premier}
     \end{eqnarray} 
for $a , b, \alpha, \beta \in \mathbb{Z}_p$. Therefore, 
the $p+1$ bases $B_0, B_1, \ldots, B_p$ constitute a complete set of MUBs in $\mathbb{C}^p$.}

\bigskip

The proof easily follows from the calculation of the modulus of $S(a - b, pb - pa + 2\alpha - 2\beta, p)$ 
from (\ref{cas d vaut 2}), (\ref{cas eeo}), (\ref{cas ooo}) and (\ref{cas oee}). As a consequence, the 
bases $B_a$, with $a = 0, 1, \ldots, p-1$, are $p$ MUBs in the sense that they satisfy (\ref{definition des mubs}) 
for any $a$, $b$, $\alpha$ and $\beta$ in the Galois field $\mathbb{F}_p$. Obviously, each of the bases 
$B_a$ (with $a = 0, 1, \ldots, p-1$) is mutually unbiased with the computational basis $B_p$. This completes 
the proof. Note that Result 6 can be proved as well from the developments in \cite{AlbouyKib}. 

As two typical examples, let us examine the cases $d=2$ and 3.

\bigskip

\noindent {\bf Example 2:} case $d=2$. 
In this case, relevant for a spin $j = 1/2$ or for a qubit, we have $q = -1$ and $a, \alpha \in \mathbb{Z}_2$. The 
matrices of the operators $v_{0a}$ are 
     \begin{eqnarray}
V_{00} = 
\pmatrix{
  0     &1   \cr
  1     &0   \cr
} \qquad 
V_{01} = 
\pmatrix{
  0     &-1  \cr
  1     &0   \cr
}.
     \end{eqnarray}
By using the notation 
	\begin{eqnarray}
\alpha := | \frac{1}{2} ,   \frac{1}{2} \rangle \qquad 
\beta  := | \frac{1}{2} , - \frac{1}{2} \rangle 
     \end{eqnarray}
familiar in quantum chemistry ($\alpha$ is a spinorbital for spin up and $\beta$ for spin down), 
the $d+1 = 3$ MUBs are 
	\begin{eqnarray}
B_{0} &:& | 0 0 \rangle =    \frac{1}{\sqrt{2}} \left( \alpha +   \beta \right) \qquad
          | 0 1 \rangle = -  \frac{1}{\sqrt{2}} \left( \alpha -   \beta \right) \\
B_{1} &:& | 1 0 \rangle =  i \frac{1}{\sqrt{2}} \left( \alpha - i \beta \right) \qquad
          | 1 1 \rangle = -i \frac{1}{\sqrt{2}} \left( \alpha + i \beta \right) \\
B_{2} &:& | 0 \rangle = \alpha \qquad
          | 1 \rangle =  \beta.    
     \label{cas d is 2 in alpha and beta}
     \end{eqnarray}

\bigskip

\noindent {\bf Example 3:} case $d=3$. 
This case corresponds to a spin $j=1$ or to a qutrit. Here, we have 
$q = \exp (i 2 \pi / 3)$ and $a, \alpha \in \mathbb{Z}_3$. The 
matrices of the operators $v_{0a}$ are 
	\begin{eqnarray}
V_{00} = 
\pmatrix{
  0     &1   &0 \cr
  0     &0   &1 \cr
  1     &0   &0 \cr
} \qquad 
V_{01} = 
\pmatrix{
  0     &q   &0   \cr
  0     &0   &q^2 \cr
  1     &0   &0   \cr
} \qquad
V_{02} = 
\pmatrix{
  0     &q^2   &0 \cr
  0     &0     &q \cr
  1     &0     &0 \cr
}.
	\end{eqnarray}
The $d+1 = 4$ MUBs read
     \begin{eqnarray}
B_{0}:  & & | 0 0 \rangle = \frac{1}{\sqrt{3}} \left(     | 0 \rangle +     | 1 \rangle + | 2 \rangle \right)  \nonumber  \\ 
        & & | 0 1 \rangle = \frac{1}{\sqrt{3}} \left( q^2 | 0 \rangle + q   | 1 \rangle + | 2 \rangle \right)  \nonumber  \\ 
        & & | 0 2 \rangle = \frac{1}{\sqrt{3}} \left( q   | 0 \rangle + q^2 | 1 \rangle + | 2 \rangle \right)             \\
B_{1}:  & & | 1 0 \rangle = \frac{1}{\sqrt{3}} \left( q   | 0 \rangle + q   | 1 \rangle + | 2 \rangle \right)  \nonumber  \\ 
        & & | 1 1 \rangle = \frac{1}{\sqrt{3}} \left(     | 0 \rangle + q^2 | 1 \rangle + | 2 \rangle \right)  \nonumber  \\ 
        & & | 1 2 \rangle = \frac{1}{\sqrt{3}} \left( q^2 | 0 \rangle +     | 1 \rangle + | 2 \rangle \right)             \\
B_{2}:  & & | 2 0 \rangle = \frac{1}{\sqrt{3}} \left( q^2 | 0 \rangle + q^2 | 1 \rangle + | 2 \rangle \right)  \nonumber  \\
        & & | 2 1 \rangle = \frac{1}{\sqrt{3}} \left( q   | 0 \rangle +     | 1 \rangle + | 2 \rangle \right)  \nonumber  \\ 
        & & | 2 2 \rangle = \frac{1}{\sqrt{3}} \left(     | 0 \rangle + q   | 1 \rangle + | 2 \rangle \right)             \\
B_{3}:  & & | 0 \rangle = | 1 , 1 \rangle \qquad | 1 \rangle = | 1 , 0 \rangle \qquad | 2 \rangle = | 1 , -1 \rangle.
     \label{cas d is 3}
     \end{eqnarray}
It should be observed that $B_0$ (respectively, $B_1$ and $B_2$) can be associated with the 
{\it vector} (respectively, {\it projective}) irreducible representations of the group $C_3$.

\subsubsection{Case $d$ power of a prime}

Different constructions of MUBs in the case where $d$ is a power of a prime 
were achieved by numerous authors from algebraical and geometrical techniques 
(see for instance \cite{WoottersFields, Kerdock, BandyoGPM5, LawrenceGPM6, Chaturvedi, Vlasov, 
Klappenecker, Gibbons, Pittenger, Durt, Archer, Aschbacher} and references therein). We want to  
show here, through an example for $d=4$, how our angular momentum approach 
can be useful for addressing this case. 

\bigskip

\noindent {\bf Example 4:} case $d=4$. 
This case corresponds to a spin $j = 3/2$. Here, we have $q = i$ and $a, \alpha \in \mathbb{Z}_4$. Equations 
(\ref{vecteur a alpha}) and (\ref{definition B_a}) can be applied to this case too. However, the resulting bases 
$B_{0}$, $B_{1}$, $B_{2}$, $B_{3}$ and $B_4$ do not constitute a complete system of MUBs 
($d=4$ is not a prime number). Nevertheless, it is possible to find $d+1 = 5$ MUBs 
because $d = 2^2$ is the power of a prime number. This can be achieved by replacing the space ${\cal E}(4)$ 
spanned by $\{ | 3/2 , m \rangle : m = 3/2, 1/2, -1/2, -3/2 \}$ by the tensor product space
${\cal E}(2) \otimes {\cal E}(2)$ spanned by the basis 
	\begin{eqnarray}
\{ \alpha \otimes \alpha, \alpha \otimes \beta, \beta \otimes \alpha, \beta \otimes \beta \}.
	\label{base pd}
	\end{eqnarray}
The space ${\cal E}(2) \otimes {\cal E}(2)$ is associated with the coupling of two spin angular momenta 
$j_1 = 1/2$ and $j_2 = 1/2$ or two qubits (in the vector $u \otimes v$, $u$ and $v$ correspond to 
$j_1$       and $j_2$, respectively). 

In addition to the basis (\ref{base pd}), it is possible to find other bases of 
${\cal E}(2) \otimes {\cal E}(2)$ which are mutually unbiased. The $d=4$ MUBs besides the 
canonical or computational basis (\ref{base pd}) can be constructed from the eigenvectors 
	\begin{eqnarray}
|a b \alpha \beta \rangle := |a \alpha \rangle \otimes |b \beta \rangle 
	\label{tensor product of vectors}
	\end{eqnarray}
of the operators 
	\begin{eqnarray}
w_{ab} := v_{0a} \otimes v_{0b}
  \label{tensor product of operators}
	\end{eqnarray} 
(the vectors $|a \alpha \rangle$ and $|b \beta \rangle$ refer 
to the two spaces ${\cal E}(2)$). As a result, we have the $d+1 = 5$ following MUBs where 
$\lambda = (1-i)/2$ and $\mu = i \lambda$.

The canonical basis:
	\begin{eqnarray}
\alpha \otimes \alpha \qquad \alpha \otimes \beta \qquad \beta \otimes \alpha \qquad \beta \otimes \beta.
	\label{canonical basis pour d4}
	\end{eqnarray}

The $w_{00}$ basis:
	\begin{eqnarray}
| 0 0 0 0 \rangle &=& \frac{1}{2} 
(\alpha \otimes \alpha + \alpha \otimes \beta + \beta \otimes \alpha + \beta \otimes \beta)  \\
| 0 0 0 1 \rangle &=& \frac{1}{2} 
(\alpha \otimes \alpha - \alpha \otimes \beta + \beta \otimes \alpha - \beta \otimes \beta)  \\
| 0 0 1 0 \rangle &=& \frac{1}{2} 
(\alpha \otimes \alpha + \alpha \otimes \beta - \beta \otimes \alpha - \beta \otimes \beta)  \\
| 0 0 1 1 \rangle &=& \frac{1}{2} 
(\alpha \otimes \alpha - \alpha \otimes \beta - \beta \otimes \alpha + \beta \otimes \beta).
	\label{w00 basis}
	\end{eqnarray}

The $w_{11}$ basis:
	\begin{eqnarray}
| 1 1 0 0 \rangle &=& \frac{1}{2} 
(\alpha \otimes \alpha + i \alpha \otimes \beta + i \beta \otimes \alpha - \beta \otimes \beta)  \\
| 1 1 0 1 \rangle &=& \frac{1}{2} 
(\alpha \otimes \alpha - i \alpha \otimes \beta + i \beta \otimes \alpha + \beta \otimes \beta)  \\
| 1 1 1 0 \rangle &=& \frac{1}{2} 
(\alpha \otimes \alpha + i \alpha \otimes \beta - i \beta \otimes \alpha + \beta \otimes \beta)  \\
| 1 1 1 1 \rangle &=& \frac{1}{2} 
(\alpha \otimes \alpha - i \alpha \otimes \beta - i \beta \otimes \alpha - \beta \otimes \beta).
	\label{w11 basis}
	\end{eqnarray}

The $w_{01}$ basis:
	\begin{eqnarray}
\lambda | 0 1 0 0 \rangle + \mu | 0 1 1 1 \rangle &=& \frac{1}{2} 
(\alpha \otimes \alpha + \alpha \otimes \beta - i \beta \otimes \alpha + i \beta \otimes \beta)  \\
\mu | 0 1 0 0 \rangle + \lambda | 0 1 1 1 \rangle &=& \frac{1}{2} 
(\alpha \otimes \alpha - \alpha \otimes \beta + i \beta \otimes \alpha + i \beta \otimes \beta)  \\
\lambda | 0 1 0 1 \rangle + \mu | 0 1 1 0 \rangle &=& \frac{1}{2} 
(\alpha \otimes \alpha - \alpha \otimes \beta - i \beta \otimes \alpha - i \beta \otimes \beta)  \\
\mu | 0 1 0 1 \rangle + \lambda | 0 1 1 0 \rangle &=& \frac{1}{2} 
(\alpha \otimes \alpha + \alpha \otimes \beta + i \beta \otimes \alpha - i \beta \otimes \beta).
	\label{w01 basis}
	\end{eqnarray}

The $w_{10}$ basis:
	\begin{eqnarray}
\lambda | 1 0 0 0 \rangle + \mu | 1 0 1 1 \rangle &=& \frac{1}{2} 
(\alpha \otimes \alpha - i \alpha \otimes \beta + \beta \otimes \alpha + i \beta \otimes \beta)  \\
\mu | 1 0 0 0 \rangle + \lambda | 1 0 1 1 \rangle &=& \frac{1}{2} 
(\alpha \otimes \alpha + i \alpha \otimes \beta - \beta \otimes \alpha + i \beta \otimes \beta)  \\
\lambda | 1 0 0 1 \rangle + \mu | 1 0 1 0 \rangle &=& \frac{1}{2} 
(\alpha \otimes \alpha + i \alpha \otimes \beta + \beta \otimes \alpha - i \beta \otimes \beta)  \\
\mu | 1 0 0 1 \rangle + \lambda | 1 0 1 0 \rangle &=& \frac{1}{2} 
(\alpha \otimes \alpha - i \alpha \otimes \beta - \beta \otimes \alpha - i \beta \otimes \beta).
	\label{w10 basis}
	\end{eqnarray}

\bigskip

It is to be noted that the vectors of the $w_{00}$ and $w_{11}$ bases are not intricated 
(i.e., each vector is the direct product of two vectors) while the vectors of the $w_{01}$ 
and $w_{10}$ bases are intricated (i.e., each vector is not the direct product 
of two vectors). To be more precise, the degree of intrication of the state 
vectors for the bases $w_{00}$, $w_{11}$, $w_{01}$ and $w_{10}$ 
can be determined in the following way. In arbitrary dimension $d$, 
let 
    \begin{eqnarray}
| \Phi \rangle = \sum_{k = 0}^{d-1} \sum_{l = 0}^{d-1} a_{kl} | k \rangle \otimes | l \rangle
    \end{eqnarray}
be a double qudit state vector. Then, it can be shown that the 
determinant of the $d \times d$ matrix $A = (a_{kl})$ satisfies 
      \begin{eqnarray}
0 \leq |\det A| \leq \frac{1}{\sqrt{d^d}}
      \end{eqnarray}
as proved in the Albouy thesis \cite{Albouythesis, Albouysubmitted}. The
case $\det A = 0$ corresponds to the absence of intrication while the case 
      \begin{eqnarray}
|\det A| = \frac{1}{\sqrt{d^d}}
      \end{eqnarray} 
corresponds to a maximal 
intrication. As an illustration,  we   obtain   that all the state vectors 
for $w_{00}$ and $w_{11}$ are not intricated and that all the state vectors 
for $w_{01}$ and $w_{10}$ are maximally intricated. 

\subsubsection{Case $d$ arbitrary} 

In the special case where $u = 1$, the generalized Gauss sum $S(1, - d + 2\alpha - 2\beta, d)$ 
can be easily calculated for $d$ arbitrary by means of the reciprocity theorem \cite{BerndtEW}
          \begin{eqnarray}
S(u, v, w) = \sqrt{ \left| \frac{w}{u} \right| } e^{i \pi [{\rm sgn} (uw) - v^2/(uw)] / 4} S(-w, -v, u).
          \label{reciprocity theorem}
          \end{eqnarray}
This leads to the following particular result.

\bigskip

\noindent {\bf Result 7.} {\it For $d$ arbitrary and $b = a \ominus 1$, one has
     \begin{eqnarray}
  \langle a \alpha | a_{-1} \beta \rangle   = 
\frac{1}{\sqrt{d}} e^{i \pi [1 - (d - 2 \alpha + 2 \beta)^2 / d] / 4} \Rightarrow 
| \langle a \alpha | a_{-1} \beta \rangle | = \frac{1}{\sqrt{d}} \qquad a_{-1} = a \ominus 1.  
     \label{d arbi un}
     \end{eqnarray}
Therefore, the three bases $B_{a \ominus 1}$, $B_a$ and $B_d$ are mutually unbiased in $\mathbb{C}^d$.}

\bigskip

This result is in agreement with a well-known result proved in many papers from quite 
distinct ways (see for instance \cite{Grassl}). We thus recover, from an approach based 
on generalized Gauss sums, that for $d$ arbitrary the minimum number of MUBs is 3. 

Another special case, viz., $u = 2$ ($\Rightarrow d \geq 3$), is worth of value. The application of the 
reciprocity theorem gives here 

\bigskip

\noindent {\bf Result 8.} {\it For $d \geq 3$ and $b = a \ominus 2$, one has 
     \begin{eqnarray}
     \langle a \alpha | a_{-2} \beta \rangle  &      =    &  \frac{1}{\sqrt{d}} \frac{1}{\sqrt{2}} 
     e^{i \pi [1 - 2 (\alpha - \beta)^2 / d] / 4} 
     [1 + e^{i \pi (-d + 2 \alpha - 2 \beta) / 2}] \nonumber \\
                                              &\Rightarrow& 
| \langle a \alpha | a_{-2} \beta \rangle | = \sqrt{\frac{2}{d}} 
\left| \cos \left[ \frac{\pi}{4} (d - 2 \alpha + 2 \beta) \right] \right| \qquad a_{-2} = a \ominus 2. 
     \label{d arbi deux }
     \end{eqnarray} 
Therefore, the bases $B_{a \ominus 2}$ and $B_a$ cannot be mutually unbiased in $\mathbb{C}^d$ 
for $d$ even with $d \geq 4$. In marked contrast, the bases $B_{a \ominus 2}$ and $B_a$ are unbiased for $d$ odd 
with $d \geq 3$ ($d$ prime or not prime).}
 
\bigskip

Going back to the Hadamard matrices, let us remark that, for $d$ arbitrary, 
if $B_a$ and $B_b$ are two MUBs associated with the Hadamard matrices $F_a$ 
and $F_b$ (respectively), then $F_a^{\dagger} F_b$ is a Hadamard matrix too. 
However, for $d$ arbitrary, if $F_a$ and $F_b$ are two Hadamard matrices 
associated with the bases $B_a$ and $B_b$ (respectively), the product 
$F_a^{\dagger} F_b$ is not in general a Hadamard matrix.

\section{Unitary group and generalized Pauli group}

     \subsection{Weyl pairs}
     
We continue with the general case where $d$ is arbitrary. 
The operator $v_{0a}$ can be expressed as 
          \begin{eqnarray}
  v_{0a} = \sum_{k=0}^{d-1} q^{ka} | k \ominus 1 \rangle \langle k |
          \Leftrightarrow 
  v_{0a} = \sum_{m = -j}^{j} q^{(j-m)a} | j , m \oplus 1 \rangle \langle j , m | 
          \label{v0a operator} 
          \end{eqnarray}
so that         
          \begin{eqnarray}
  v_{0a} | k \rangle = q^{ka} | k \ominus 1 \rangle 
          \Leftrightarrow 
  v_{0a} | j , m \rangle = q^{(j-m)a} | j , m \oplus 1 \rangle          
          \label{action of v0a on k} 
          \end{eqnarray}
where $q = \exp(2 \pi i / d)$. The operators $x$ (the flip or shift operator) and $z$ (the clock operator), 
used in quantum information and quantum computation 
(see for instance \cite{geometricalanalysisofPdHS, geometricalanalysisofPdPB}), 
can be derived from the generic operator $v_{0a}$ as follows 
 	\begin{eqnarray}
x := v_{00} \qquad z := \left( v_{00} \right) ^{\dagger} v_{01}. 
	\label{definition of x and z}
	\end{eqnarray}
Therefore, we get
 	\begin{eqnarray}
x = \sum_{k=0}^{d-1} | k \ominus 1 \rangle \langle k | = 
| d       - 1 \rangle \langle   0 | +
|           0 \rangle \langle   1 | +
\ldots +
| d       - 2 \rangle \langle d-1 |
	\label{expression of x}
	\end{eqnarray}
and
 	\begin{eqnarray}	
z = \sum_{k=0}^{d-1} q^k | k \rangle \langle k | = 
        | 0   \rangle \langle   0 | +
      q | 1   \rangle \langle   1 | +
\ldots +
q^{d-1} | d-1 \rangle \langle d-1 |.
	\label{expression of z}
	\end{eqnarray}
The action of $x$ and $z$ on the basis $B_d$ of ${\cal E}(d)$ is given 
by the ladder relation 
 	\begin{eqnarray}	
x | k \rangle = | k \ominus 1 \rangle \Leftrightarrow 
x |j , m \rangle = \left( 1 - \delta_{m,j} \right) |j , m+1 \rangle + \delta_{m,j} |j , -j \rangle        
	\label{action of x on Bd}
	\end{eqnarray}
and the phase relation 
          \begin{eqnarray}
z | k \rangle = q^{k} | k \rangle \Leftrightarrow
z | j,m \rangle = q^{j-m} | j,m \rangle.   
          \label{action of z on Bd} 
          \end{eqnarray}
Alternatively, the action of $x$ and $z$ on any basis $B_a$ ($a = 0, 1, \ldots, d-1$) of ${\cal E}(d)$ 
reads
 	\begin{eqnarray}	
x | a \alpha \rangle = q^{(d-1)a / 2 - \alpha} | a \alpha_a \rangle \qquad \alpha_a = \alpha \oplus a 
\qquad \Rightarrow \qquad 
x | 0 \alpha \rangle = q^{           - \alpha} | 0 \alpha   \rangle       
	\label{action of x on Ba}
	\end{eqnarray}
and  
          \begin{eqnarray}
z | a \alpha \rangle = q^{-1} | a \alpha_{-1} \rangle \qquad \alpha_{-1} = \alpha \ominus 1.   
          \label{action of z on Ba} 
          \end{eqnarray} 
Equations (\ref{action of x on Bd}) and (\ref{action of z on Bd}), on one side, and 
     Eqs.~(\ref{action of x on Ba}) and (\ref{action of z on Ba}), on the other side, 
show that the flip or clock character for $x$ and $z$ is basis-dependent. The 
relationship between $x$ and $z$ can be understood via the following 

\bigskip

\noindent {\bf Result 9.} {\it The unitary operators $x$ and $z$ are cyclic and $q$-commute: 
     \begin{eqnarray}
x^d = z^d = I \qquad x z - q z x = 0. 
     \label{paire de W}
     \end{eqnarray}
They are connected by 
          \begin{eqnarray}
x = f^{\dagger} z f \Leftrightarrow z = f x f^{\dagger} 
          \label{x-z connexion} 
          \end{eqnarray}
where the Fourier operator  
          \begin{eqnarray}
f := \frac{1}{\sqrt{d}} \sum_{k = 0}^{d-1} \sum_{k' = 0}^{d-1} q^{- k k'} 
  | k \rangle  \langle k' | 
          \end{eqnarray} 
is unitary and satisfies 
          \begin{eqnarray}
f^4 = 1. 
          \end{eqnarray}   
The operators $x$ and $z$ are isospectral operators with the 
common spectrum $\{ 1, q, \ldots, q^{d-1} \}$.}

\bigskip

A direct proof of Result 9 can be obtained by switching to the matrices 
        \begin{eqnarray}
X = \sum_{k=0}^{d-1}     E_{k \ominus 1 , k} = 
\pmatrix{
0                    &      1 &      0  & \ldots &       0 \cr
0                    &      0 &      1  & \ldots &       0 \cr
\vdots               & \vdots & \vdots  & \ldots &  \vdots \cr
0                    &      0 &      0  & \ldots &       1 \cr
1                    &      0 &      0  & \ldots &       0 \cr
}
        \end{eqnarray}
        \begin{eqnarray}
Z = \sum_{k=0}^{d-1} q^k E_{k , k} = 
\pmatrix{
1                    &      0 &      0    & \ldots &       0       \cr
0                    &      q &      0    & \ldots &       0       \cr
0                    &      0 &      q^2  & \ldots &       0       \cr
\vdots               & \vdots & \vdots    & \ldots &  \vdots       \cr
0                    &      0 &      0    & \ldots &       q^{d-1} \cr
}
        \label{definition of X and Z en E}
        \end{eqnarray}
of the operators $x$ and $z$, in the basis $B_d$ (cf.~(\ref{action of z on Bd}) 
and (\ref{action of x on Ba})). Let $F$ be the matrix of the linear operator 
$f$ in the basis $B_d$. The reduction by means of $F$ of the endomorphism 
associated with the matrix $X$ yields the matrix $Z$. In other words, the 
diagonalization of $X$ can be achieved with the help of the matrix $F$ via 
$Z = F X F^{\dagger}$. Note that the matrix $F$ is connected to $F_0$ by 
          \begin{eqnarray}
F = (F_0 S)^{\dagger} \qquad 
S:= \sum_{\beta = 0}^{d-1} q^{\beta} E_{\beta, d-\beta}
          \label{connexion H0-F}
          \end{eqnarray}
where $S$ acts as a pseudopermutation.

In view of (\ref{paire de W}), the pair $(x , z)$ is called a Weyl pair. Weyl pairs were originally 
introduced in finite quantum mechanics \cite{Weyl} and used for the construction of unitary bases 
in finite-dimensional Hilbert spaces \cite{Schwinger}. It should be noted that matrices of type $X$ 
and $Z$ were introduced long time ago by Sylvester \cite{Sylvester} in order to solve the matrix 
equation $PX = XQ$; in addition, such matrices were used by Morris \cite{Morris} to define 
generalized Clifford algebras in connection with quaternion algebras and division rings. Besides the 
Weyl pair $(x , z)$, other pairs can be formed with the operators $v_{0a}$ and $z$. Indeed, any 
operator $v_{0a}$ ($a \in \mathbb{Z}_d$) can be generated from $x$ and $z$ since 
     \begin{eqnarray}
v_{0a} = x z^a.
     \label{v0a en x et z}
     \end{eqnarray}
Thus, Eq.~(\ref{paire de W}) can be generalized as 
     \begin{eqnarray}
e^{-i \pi (d-1)a} (v_{0a})^d = z^d = I \qquad v_{0a} z - q z v_{0a} = 0. 
     \label{pseudo paire de W}
     \end{eqnarray}
Therefore, the pair $(v_{0a} , z)$ is a Weyl pair for $(d-1)a$ even. 

     \subsection{Generalized Pauli matrices}
     
For $d=2$ the $q$-commutation relation of $x$ and $z$ reduces 
to an anticommutation relation. In fact, Eq.~(\ref{paire de W}) 
with  $d=2$ can be particularized to the relations
     \begin{eqnarray}
x^2 = z^2 = I \qquad x z + z x = 0 
     \label{paire de W pour d2}
     \end{eqnarray}
which are reminiscent of relations satisfied by the Pauli matrices. Hence, we understand that the 
matrices $X$ and $Z$ for $d$ arbitrary can be used as an integrity basis for producing generalized 
Pauli matrices \cite{Ivanovic, WoottersFields, Kerdock, BandyoGPM5, LawrenceGPM6, Vlasov, Klappenecker, 
Gibbons, Pittenger, Balian, PateraZassenhaus, Galetti, Knill, Gott, Pittenger2000, Kitaev, Bartlett, 
Klimov, Romero}. Let us develop this point. 

For $d$ arbitrary, we define the operators 
     \begin{eqnarray}
u_{ab} = x^a z^b \qquad a, b \in \mathbb{Z}_d.
     \label{generalized Pauli operators}
     \end{eqnarray}
The operators $u_{ab}$ shall be referred as generalized Pauli operators and their matrices as generalized 
Pauli matrices. They satisfy the ladder-phase relation
     \begin{eqnarray}
u_{ab} |  k  \rangle = q^{kb}     |   k   \ominus a \rangle \Leftrightarrow 
u_{ab} | j,m \rangle = q^{(j-m)b} | j , m \oplus  a \rangle
     \label{ladder phase relation}
     \end{eqnarray}
from which we can derive the following result. 

\bigskip

\noindent {\bf Result 10.} {\it The $d^2$ operators $u_{ab}$, with $a, b \in \mathbb{Z}_d$, are unitary 
and obey the multiplication rule 
          \begin{eqnarray}
u_{ab} u_{a'b'} = q^{-ba'} u_{a'' b''}  \qquad 
a'' := a \oplus a'  \qquad 
b'' := b \oplus b'.
          \label{produit uu}
          \end{eqnarray}
Therefore, the commutator and the 
anticommutator of $u_{ab}$ and $u_{a'b'}$ are given by 
          \begin{eqnarray}
[u_{ab} , u_{a'b'}]_{\pm} = \left( q^{-ba'} \pm q^{-ab'} \right) u_{a'' b''}  \qquad 
a'' := a \oplus a'  \qquad b'' := b \oplus b'. 
          \label{com anti-com}
          \end{eqnarray}
Furthermore, they are orthogonal with respect to the Hilbert-Schmidt inner product 
          \begin{eqnarray}
 {\rm Tr}_{{\cal E}(d)} \left[ (u_{ab})^{\dagger} u_{a'b'} \right] = 
 d \>
 \delta_{a,a'} \> 
 \delta_{b,b'} 
          \label{trace de uu}
          \end{eqnarray}
where the trace is taken on the $d$-dimensional space ${\cal E}(d)$.}

\bigskip
        
As a corollary of Result 10, we have 
          \begin{eqnarray}
[u_{ab} , u_{a'b'}]_{-} = 0  \Leftrightarrow ab' \ominus ba' = 0
          \label{com}
          \end{eqnarray}
and
          \begin{eqnarray}
[u_{ab} , u_{a'b'}]_{+} = 0  \Leftrightarrow ab' \ominus ba' = \frac{1}{2} d.
          \label{anti-com en ab}
          \end{eqnarray}
This yields two consequences. First, Eq.~(\ref{anti-com en ab}) shows that all anticommutators $[u_{ab} , u_{a'b'}]_{+}$ are
different from 0 if $d$ is an odd integer. Second, from  Eq.~(\ref{com}) we have the important result that, 
for $d$ arbitrary, each of the three disjoint sets 
          \begin{eqnarray}
e_{      0 \bullet} &:=& \{ u_{0a} =     z^a : a = 1, 2, \ldots, d-1 \}  \\
e_{\bullet \bullet} &:=& \{ u_{aa} = x^a z^a : a = 1, 2, \ldots, d-1 \}  \\
e_{\bullet       0} &:=& \{ u_{a0} = x^a     : a = 1, 2, \ldots, d-1 \} 
          \label{les 3 e}
          \end{eqnarray}
consist of $d-1$ mutually commuting operators. The three sets  
$e_{      0 \bullet}$, 
$e_{\bullet \bullet}$ and  
$e_{\bullet       0}$ are associated with three MUBs. This is in agreement with the fact that the bases $B_0$, 
$B_1$ and $B_d$ are three MUBs for $d$ arbitrary 
($v_{00} = x  \in e_{\bullet       0}$, 
 $v_{01} = xz \in e_{\bullet \bullet}$ and 
 $          z \in e_{0       \bullet}$ are associated with $B_0$, $B_1$ and $B_d$, respectively). 

By way of illustration, let us give the matrices in the 
basis $B_d$ of the operators $u_{ab}$ for $d=2$, 3 and 4.

\bigskip

\noindent {\bf Example 5:} case $d=2$. 
For $d = 2 \Leftrightarrow j = 1/2$ ($\Rightarrow q = -1$), the matrices in the two sets 
          \begin{eqnarray}
E_{0} &:=& \{ I_2 = X^0 Z^0, X = X^1 Z^0 \equiv V_{00} \}  \\
E_{1} &:=& \{ Z   = X^0 Z^1, Y = X^1 Z^1 \equiv V_{01} \}  
          \label{les 2 E de d2}
          \end{eqnarray}
corresponding to the four operators $u_{ab}$ are 
     \begin{eqnarray}
I_2 =  
\pmatrix{
  1     &0   \cr
  0     &1   \cr
} \qquad 
X = 
\pmatrix{
  0     &1   \cr
  1     &0   \cr
} \qquad
Z = 
\pmatrix{
  1     &0   \cr
  0     &-1  \cr
} \qquad 
Y = 
\pmatrix{
  0     &-1  \cr
  1     &0   \cr
}.
     \end{eqnarray}
In terms of the usual (Hermitian and unitary) Pauli matrices 
$\sigma_x$, $\sigma_y$ and $\sigma_z$, we have 
\begin{eqnarray}
X =     \sigma_x \qquad   
Y = - i \sigma_y \qquad 
Z =     \sigma_z. 
\end{eqnarray}
The matrices $X$, $Y$ 
and $Z$ are thus identical to the Pauli matrices up to a phase 
factor for $Y$. This phase factor is the price one has to pay 
in order to get a systematic generalization of Pauli matrices 
in arbitrary dimension.

\bigskip

\noindent {\bf Example 6:} case $d=3$. 
For $d = 3 \Leftrightarrow j = 1$ ($\Rightarrow q = \exp(i 2 \pi / 3)$), 
the matrices in the three sets 
          \begin{eqnarray}
E_{0} &:=& \{ X^0 Z^0, X^1 Z^0 \equiv V_{00}, X^2 Z^0 \}  \\
E_{1} &:=& \{ X^0 Z^1, X^1 Z^1 \equiv V_{01}, X^2 Z^1 \}  \\
E_{2} &:=& \{ X^0 Z^2, X^1 Z^2 \equiv V_{02}, X^2 Z^2 \}
          \label{les 3 E de d3}
          \end{eqnarray}
corresponding to the nine operators $u_{ab}$ are 
    \begin{eqnarray}
I_3 = 
\pmatrix{
  1     &0     &0   \cr
  0     &1     &0   \cr
  0     &0     &1   \cr
} \qquad 
X = 
\pmatrix{
  0     &1     &0   \cr
  0     &0     &1   \cr
  1     &0     &0   \cr
} \qquad 
X^2 = 
\pmatrix{
  0     &0     &1   \cr
  1     &0     &0   \cr
  0     &1     &0   \cr
}
     \end{eqnarray}
     \begin{eqnarray}
Z = 
\pmatrix{
  1     &0     &0     \cr
  0     &q     &0     \cr
  0     &0     &q^2   \cr
} \qquad 
X Z = 
\pmatrix{
  0     &q     &0     \cr
  0     &0     &q^2   \cr
  1     &0     &0     \cr
} \qquad
X^2 Z =  
\pmatrix{
  0     &0     &q^2     \cr
  1     &0     &0       \cr
  0     &q     &0       \cr
}
     \end{eqnarray}
     \begin{eqnarray}
Z^2 = 
\pmatrix{
  1     &0       &0   \cr
  0     &q^2     &0   \cr
  0     &0       &q   \cr
} \qquad 
X Z^2 =  
\pmatrix{
  0     &q^2     &0     \cr
  0     &0       &q     \cr
  1     &0       &0     \cr
} \qquad     
X^2 Z^2 = 
\pmatrix{
  0     &0       &q     \cr
  1     &0       &0     \cr
  0     &q^2     &0     \cr
}.
     \end{eqnarray}
These generalized Pauli matrices differ from the Gell-Mann matrices and Okubo matrices used for 
$SU(3)$ in  particle physics with three flavors of quarks \cite{Ne'eman, Gell-Mann, Okubo}. They 
constitute a natural extension based on Weyl pairs of the Pauli matrices in dimension $d = 3$. 

\bigskip

\noindent {\bf Example 7:} case $d=4$. 
For $d = 4 \Leftrightarrow j = 3/2$ ($\Rightarrow q = i)$, 
the matrices in the four sets 
          \begin{eqnarray}
E_{0} &:=& \{ X^0 Z^0, X^1 Z^0 \equiv V_{00}, X^2 Z^0, X^3 Z^0 \}  \\
E_{1} &:=& \{ X^0 Z^1, X^1 Z^1 \equiv V_{01}, X^2 Z^1, X^3 Z^1 \}  \\
E_{2} &:=& \{ X^0 Z^2, X^1 Z^2 \equiv V_{02}, X^2 Z^2, X^3 Z^2 \}  \\
E_{3} &:=& \{ X^0 Z^3, X^1 Z^3 \equiv V_{03}, X^2 Z^3, X^3 Z^3 \}
          \label{les 4 E de d4}
          \end{eqnarray}
corresponding to the 16 operators $u_{ab}$ are
      \begin{eqnarray}
I_4 = 
\pmatrix{
  1     &0     &0   &0 \cr
  0     &1     &0   &0 \cr
  0     &0     &1   &0 \cr
  0     &0     &0   &1 \cr
} \qquad     
X = 
\pmatrix{
  0     &1     &0   &0 \cr
  0     &0     &1   &0 \cr
  0     &0     &0   &1 \cr
  1     &0     &0   &0 \cr
}
       \end{eqnarray}
       \begin{eqnarray}
X^2 = 
\pmatrix{
  0     &0     &1   &0 \cr
  0     &0     &0   &1 \cr
  1     &0     &0   &0 \cr
  0     &1     &0   &0 \cr
} \qquad 
X^3 = 
\pmatrix{
  0     &0     &0   &1 \cr
  1     &0     &0   &0 \cr
  0     &1     &0   &0 \cr
  0     &0     &1   &0 \cr
}
      \end{eqnarray}
      \begin{eqnarray}
Z = 
\pmatrix{
  1     &0     &0    &0  \cr
  0     &i     &0    &0  \cr
  0     &0     &-1   &0  \cr
  1     &0     &0    &-i \cr
} \qquad 
X Z = 
\pmatrix{
  0     &i     &0    &0  \cr
  0     &0     &-1   &0  \cr
  0     &0     &0    &-i \cr
  1     &0     &0    &0  \cr
} 
      \end{eqnarray}
      \begin{eqnarray}
X^2 Z = 
\pmatrix{
  0     &0     &-1   &0  \cr
  0     &0     &0    &-i \cr
  1     &0     &0    &0  \cr
  0     &i     &1    &0  \cr
} \qquad  
X^3 Z = 
\pmatrix{
  0     &0     &0    &-i \cr
  1     &0     &0    &0  \cr
  0     &i     &0    &0  \cr
  0     &0     &-1   &0  \cr
} 
      \end{eqnarray}
      \begin{eqnarray}
Z^2 = 
\pmatrix{
  1     &0      &0   &0  \cr
  0     &-1     &0   &0  \cr
  0     &0      &1   &0  \cr
  0     &0      &0   &-1 \cr
} \qquad 
X Z^2 = 
\pmatrix{
  0     &-1    &0   &0  \cr
  0     &0     &1   &0  \cr
  0     &0     &0   &-1 \cr
  1     &0     &0   &0  \cr
}
      \end{eqnarray}
      \begin{eqnarray}
X^2 Z^2 = 
\pmatrix{
  0     &0     &1   &0  \cr
  0     &0     &0   &-1 \cr
  1     &0     &0   &0  \cr
  0     &-1    &0   &0  \cr
} \qquad 
X^3 Z^2 = 
\pmatrix{
  0     &0      &0   &-1  \cr
  1     &0      &0   &0   \cr
  0     &-1     &0   &0   \cr
  0     &0      &1   &0   \cr
} 
      \end{eqnarray}
      \begin{eqnarray}
Z^3 = 
\pmatrix{
  1     &0     &0   &0 \cr
  0     &-i    &0   &0 \cr
  0     &0     &-1  &0 \cr
  0     &0     &0   &i \cr
} \qquad   
X Z^3 = 
\pmatrix{
  0     &-i     &0    &0 \cr
  0     &0      &-1   &0 \cr
  0     &0      &0    &i \cr
  1     &0      &0    &0 \cr
} 
     \end{eqnarray}
     \begin{eqnarray}
X^2 Z^3 = 
\pmatrix{
  0     &0      &-1   &0 \cr
  0     &0      &0    &i \cr
  1     &0      &0    &0 \cr
  0     &-i     &0    &0 \cr
} \qquad
X^3 Z^3 = 
\pmatrix{
  0     &0     &0    &i \cr
  1     &0     &0    &0 \cr
  0     &-i    &0    &0 \cr
  0     &0     &-1   &0 \cr
}.
     \end{eqnarray}
These generalized Pauli matrices are linear combinations of the generators of the chain 
$SU(4) \supset SU(3) \supset SU(2)$ in particle physics with four flavors of quarks 
\cite{GIM, Georgi-Glashow, Moffat}. 

\bigskip

For $d$ arbitrary, the generalized Pauli matrices arising from (\ref{generalized Pauli operators}) 
are different from the generalized Gell-Mann $\lambda$ matrices introduced in \cite{deAzcarraga}. The 
generalized $\lambda$ matrices are Hermitian and adapted to the 
chain of groups $SU(d) \supset SU(d-1) \supset \ldots \supset SU(2)$ while the matrices $X^a Z^b$ 
are unitary and closely connected to cyclic symmetry. Indeed, for $d$ arbitrary, each of the $d$ sets 
          \begin{eqnarray}
E_{b} := \{ X^a Z^b : a = 0, 1, \ldots, d-1 \} \qquad b = 0, 1, \ldots, d-1
          \label{les d E de dd}
          \end{eqnarray}
is associated with an irreducible representation of the cyclic group $C_d$. More precisely, the 
one-dimensional irreducible representation of $C_d$ associated with $E_{b}$ is obtained by listing 
the nonzero matrix elements of any matrix of $E_{b}$, column by column from left to right. In this way, 
we obtain the $d$ irreducible representations of $C_d$. This relationship between $d$-dimensional Pauli 
matrices and irreducible representations of $C_d$ are clearly emphasized by the examples given above 
for $d=2$, 3 and 4.

     \subsection{Pauli basis for the unitary group}

Two consequences follow from (\ref{trace de uu}). (i) The Hilbert-Schmidt relation (\ref{trace de uu}) 
in the Hilbert space $\mathbb{C}^{d^2}$ shows that the $d^2$ operators $u_{ab}$ are pairwise 
orthogonal operators. Thus, they can serve as a basis for developing any operator acting on 
${\cal E}(d)$. (ii) The commutator in (\ref{com anti-com}) defines the Lie bracket of a 
$d^2$-dimensional Lie algebra generated by the set $\{ u_{ab} : a,b = 0, 1, \ldots, d-1 \}$. 
This algebra can be identified to the Lie algebra $u(d)$ of the unitary group $U(d)$. The 
subset $\{ u_{ab} : a,b = 0, 1, \ldots, d-1 \} \setminus \{ u_{00} \}$ then spans the Lie 
algebra $su(d)$ of the special unitary group $SU(d)$. In other words, the Weyl pair $(X,Z)$, 
consisting of the generalized Pauli matrices $X$ and $Z$ in dimension $d$, form an integrity 
basis for $u(d)$. More specifically, the two following results hold. 

\bigskip

\noindent {\bf Result 11.} {\it The set $\{ X^a Z^b : a,b = 0, 1, \ldots, d-1 \}$ forms 
a basis for the Lie algebra $u(d)$ of the unitary group $U(d)$. The Lie 
brackets of $u(d)$ in such a basis (denoted as the Pauli basis) are 
     \begin{eqnarray}
[ X^a Z^b , X^{e} Z^{f} ]_- = \sum_{i=0}^{d-1} \sum_{j=0}^{d-1} (ab,ef;ij) X^{i} Z^{j} 
     \end{eqnarray}
with the structure constants  
      \begin{eqnarray}
(ab,ef;ij) = \delta(i, a \oplus e) 
             \delta(j, b \oplus f)
		 \left( q^{- be} - q^{- af} \right)   
     \end{eqnarray}
where $a, b, e, f, i, j \in \mathbb{Z}_d$. The structure constants 
$(ab,ef;ij)$ with $i=a \oplus e$ and $j=b \oplus f$ are cyclotomic 
polynomials associated with $d$. They vanish for $af \ominus be = 0$.}   

\bigskip

\noindent {\bf Result 12.} {\it For $d=p$, with $p$ a prime integer, the Lie algebra 
$su(p)$ of the special unitary group $SU(p)$ can be decomposed 
into a direct sum of $p+1$ abelian subalgebras of dimension $p-1$, i.e.
                  \begin{eqnarray}
{su}(p) \simeq 
{ v}_0     \uplus 
{ v}_1     \uplus 
\ldots     \uplus      
{ v}_{p}     
                  \label{decompo de su(p)}
                  \end{eqnarray}
where each of the $p+1$ subalgebras ${ v}_0, { v}_1, \ldots, { v}_p$ is 
a Cartan subalgebra generated by a set of $p - 1$ commuting matrices. The 
various sets are                 
           \begin{eqnarray}  
{\cal V}_0       &:=      &  \{ X^0 Z^1    , X^0 Z^2    , X^0Z^3,     \ldots, X^0Z^{p-2},     X^0     Z^{p-1} \} 
    \label{ensemble V0}
  \\                   
{\cal V}_1       &:=      &  \{ X^1 Z^0    , X^2 Z^0    , X^3Z^0,     \ldots, X^{p-2}Z^0,     X^{p-1}     Z^0 \}   
    \label{ensemble V1}
  \\
{\cal V}_2       &:=      &  \{ X^1 Z^1    , X^2 Z^2    , X^3Z^3,     \ldots, X^{p-2}Z^{p-2}, X^{p-1} Z^{p-1} \}  
  \\
{\cal V}_3       &:=      &  \{ X^1 Z^2    , X^2 Z^4    , X^3Z^6,     \ldots, X^{p-2}Z^{p-4}, X^{p-1} Z^{p-2} \} 
  \\
                 &\vdots  & 
  \\
 {\cal V}_{p-1}  &:=      &  \{ X^1 Z^{p-2}, X^2 Z^{p-4}, X^3Z^{p-6}, \ldots, X^{p-2}Z^4,      X^{p-1}  Z^2   \} 
  \\  
 {\cal V}_{p}    &:=      &  \{ X^1 Z^{p-1}, X^2 Z^{p-2}, X^3Z^{p-3}, \ldots, X^{p-2}Z^2,      X^{p-1}    Z^1 \} 
    \label{ensemble Vp} 
          \end{eqnarray} 
for ${ v}_0, { v}_1, \ldots, { v}_{p}$, respectively.} 

\bigskip

\noindent {\bf Example 8:} $p = 7 \Leftrightarrow j = 3$. Equations (\ref{ensemble V0})-(\ref{ensemble Vp}) give
           \begin{eqnarray}  	   
{\cal V}_0       &=      &  \{ ( 01 ), ( 02 ), ( 03 ), ( 04 ), ( 05 ), ( 06 ) \} 
  \\	   	   
{\cal V}_1       &=      &  \{ ( 10 ), ( 20 ), ( 30 ), ( 40 ), ( 50 ), ( 60 ) \} 
  \\
{\cal V}_2      &=      &  \{ ( 11 ), ( 22 ), ( 33 ), ( 44 ), ( 55 ), ( 66 )  \} 
  \\
{\cal V}_3      &=      &  \{ ( 12 ), ( 24 ), ( 36 ), ( 41 ), ( 53 ), ( 65 )  \} 
  \\
{\cal V}_4      &=      &  \{ ( 13 ), ( 26 ), ( 32 ), ( 45 ), ( 51 ), ( 64 )  \} 
  \\
{\cal V}_5      &=      &  \{ ( 14 ), ( 21 ), ( 35 ), ( 42 ), ( 56 ), ( 63 )  \} 
  \\
{\cal V}_6      &=      &  \{ ( 15 ), ( 23 ), ( 31 ), ( 46 ), ( 54 ), ( 62 )  \} 
  \\
{\cal V}_7      &=      &  \{ ( 16 ), ( 25 ), ( 34 ), ( 43 ), ( 52 ), ( 61 )  \} 
          \end{eqnarray} 
where $(ab)$ is used as an abbreviation of $X^a Z^b$.

\bigskip

Result 12 can be extended to the case where $d = p^e$ 
with $p$ a prime integer and $e$ a positive integer: there 
exists a decomposition of $su(p^e)$ into $p^e+1$ abelian 
subalgebras of dimension $p^e - 1$. In order to make 
this point clear, we start with a counterexample.

\bigskip

\noindent {\bf Counterexample:} $d=4  \Leftrightarrow j = 3/2$ 
($\Rightarrow a, b = 0,1, 2, 3$). In this case, Result 11 is valid but 
Result 12 does not apply. Indeed, the 16 unitary operators $u_{ab}$ corresponding to 
\begin{eqnarray} 
ab = 01, 02, 03, 10, 20, 30, 11, 22, 33, 12, 13, 21, 23, 31, 32, 00
\label{les ab en dim 4}
\end{eqnarray}
are linearly independent and span the Lie algebra of $U(4)$ but they give only three 
disjoint sets, viz., $\{ (01), (02), (03) \}$, $\{ (10), (20), (30) \}$ and  
$\{ (11), (22), (33) \}$, containing each 3 commuting operators, where 
here again $(ab)$ stands for $X^a Z^b$. However, it is not possible to
partition the set (\ref{les ab en dim 4}) in order to get a decomposition
similar to (\ref{decompo de su(p)}). Nevertheless, it is possible to find 
another basis of $u(4)$ which can be partitioned in a way yielding a
decompostion similar to (\ref{decompo de su(p)}). This can be achieved by 
working with tensorial products of the matrices $X^a Z^b$ corresponding 
to $p=2$. In this respect, let us consider the product $u_{a_1b_1} \otimes u_{a_2b_2}$, 
where $u_{a_ib_i}$ with $i = 1,2$ are Pauli operators for $p=2$. Then, by using
the abbreviation $(a_1b_1a_2b_2)$ for $u_{a_1b_1} \otimes u_{a_2b_2}$ or 
$X^{a_1} Z^{b_1} \otimes X^{a_2} Z^{b_2}$, it can be checked that the five disjoint 
sets
\begin{eqnarray} 
\{ (1011), (1101), (0110) \} 
\label{set1} \\
\{ (1110), (1001), (0111) \} 
\label{set2} \\
\{ (1010), (1000), (0010) \} 
\label{set3} \\
\{ (1111), (1100), (0011) \} 
\label{set4} \\
\{ (0101), (0100), (0001) \} 
\label{set5}
\end{eqnarray} 
consist each of 3 commuting unitary operators and that the Lie algebra $su(4)$ 
is spanned by the union of the 5 sets. It is to be emphasized that the 15 
operators (\ref{set1})-(\ref{set5}) are underlaid by the geometry of the generalized 
quadrangle of order 2 \cite{PlanatGPM11}. In this geometry, the five sets given 
by (\ref{set1})-(\ref{set5}) correspond to a spread of this quadrangle, i.e., to a 
set of 5 pairwise skew lines \cite{PlanatGPM11}.

\bigskip

The considerations of the counterexample can be generalized to 
$d := d_1 d_2 \ldots d_e$, $e$ being an integer greater or equal to
$2$. Let us define 
   \begin{eqnarray} 
u_{AB} := u_{a_1b_1} \otimes u_{a_2b_2} \otimes \ldots \otimes u_{a_eb_e}
\quad A := a_1, a_2, \ldots, a_e
\quad B := b_1, b_2, \ldots, b_e
   \label{u_AB}
   \end{eqnarray} 
where $u_{a_1b_1}, u_{a_2b_2}, \ldots,  u_{a_eb_e}$ are generalized Pauli operators 
corresponding to the dimensions $d_1, d_2, \ldots, d_e$ respectively. In addition, 
let $q_1, q_2, \ldots, q_e$ be the $q$-factor associated 
with $d_1, d_2, \ldots, d_e$ respectively ($q_j := \exp (2 \pi i / d_j)$). Then, 
Results 10, 11 and 12 can be generalized as follows.

\bigskip

\noindent {\bf Result 13.} {\it The operators $u_{AB}$ are unitary and satisfy the 
orthogonality relation
          \begin{eqnarray}
 {\rm Tr}_{{\cal E}(d_1 d_2 \ldots d_e)} \left[ \left( u_{AB} \right)^{\dagger} u_{A'B'} \right] = 
 d_1 d_2 \ldots d_e \>
 \delta_{A,A'} \> 
 \delta_{B,B'} 
          \label{trace de uABuA'B'}
          \end{eqnarray}
where 
         \begin{eqnarray}
\delta_{A,A'} := \delta_{a_1,a_1'} \delta_{a_2,a_2'} \ldots \delta_{a_e,a_e'} \qquad 
\delta_{B,B'} := \delta_{b_1,b_1'} \delta_{b_2,b_2'} \ldots \delta_{b_e,b_e'}. 
         \label{delta de AA' et BB'}
          \end{eqnarray}
The commutator 
                $[u_{AB} , u_{A'B'}]_-$ and the 
anti-commutator $[u_{AB} , u_{A'B'}]_+$ of $u_{AB}$ and $u_{A'B'}$ are given by 
          \begin{eqnarray}
[u_{AB} , u_{A'B'}]_{\mp} = \left( \prod_{j=1}^e q_j^{-b_ja_j'} \mp 
                                   \prod_{j=1}^e q_j^{-a_jb_j'} \right) u_{A'' B''}  
          \label{com anti-com en AB}
          \end{eqnarray}
with
          \begin{eqnarray}
A'' := a_1 \oplus a_1', a_2 \oplus a_2', \ldots, a_e \oplus a_e' \qquad  
B'' := b_1 \oplus b_1', b_2 \oplus b_2', \ldots, b_e \oplus b_e'.   
          \label{A''B''}
          \end{eqnarray}
The set 
$\{ u_{AB} : A, B \in \mathbb{Z}_{d_1} \otimes \mathbb{Z}_{d_2} \otimes \ldots \otimes \mathbb{Z}_{d_e} \}$ 
of the $d_1^2 d_2^2 \ldots d_e^2$ 
unitary operators $u_{AB}$ form a basis for the Lie algebra $u(d_1 d_2 \ldots d_e)$ 
of the group $U(d_1 d_2 \ldots d_e)$.} 

\bigskip

The operators $u_{AB}$ may be called generalized Dirac 
operators since the ordinary Dirac operators correspond to 
specific $u_{a_1b_1} \otimes u_{a_2b_2}$ for $d_1 = d_2 = 2$.

In the special case where $d_1 = d_2 = \ldots = d_e = p$ with $p$ a prime integer 
(or equivalently $d = p^e$), we have $[u_{AB} , u_{A'B'}]_{-} = 0$ if and only if 
          \begin{eqnarray}
\sum_{j=1}^e a_jb_j' \ominus b_ja_j' = 0 
          \end{eqnarray}
and $[u_{AB} , u_{A'B'}]_{+} = 0$ if and only if 
          \begin{eqnarray}
\sum_{j=1}^e a_jb_j' \ominus b_ja_j' = \frac{1}{2} p 
          \label{anti-com en AB}
          \end{eqnarray}
so that there are vanishing anti-commutators only if $p = 2$. The commutation relations 
given by (\ref{com anti-com en AB})-(\ref{A''B''}) can be transcribed in terms of 
Lagrangian submodules \cite{Albouythesis, Albouysubmodule}. For $d = p^e$, there exists a 
decomposition of the set $\{u_{AB} : A, B \in \mathbb{Z}_{p}^{\otimes e}\} \setminus \{I\}$ 
that corresponds to a decomposition of the Lie algebra $su(p^e)$ into $p^e +1$ abelian 
subalgebras of dimension $p^e - 1$ 
\cite{KibJPhysA, PateraZassenhaus, KKU, autresdecomp1, autresdecomp2, autresdecomp3}. 

     \subsection{Generalized Pauli group}

Let us define the $d^3$ operators 
          \begin{eqnarray}
w_{abc} := q^a x^b z^c = q^a u_{bc} \qquad a, b, c \in \mathbb{Z}_d. 
          \end{eqnarray} 
The action of $w_{abc}$ on the Hilbert space ${\cal E} (d)$ is described by 
     \begin{eqnarray}
w_{abc} |  k   \rangle = q^{a + kc}     |    k \ominus b \rangle \Leftrightarrow 
w_{abc} |j , m \rangle = q^{a + (j-m)c} |j , m \oplus  b \rangle. 
     \label{action de wabc sur jm} 
     \end{eqnarray}
The operators $w_{abc}$ are unitary and satisfy 
          \begin{eqnarray}
{\rm Tr}_{{\cal E}(d)} \left[ (w_{abc})^{\dagger} w_{a'b'c'} \right] = 
 q^{a' - a} \> d \>
 \delta_{b,b'} \> 
 \delta_{c,c'}
          \label{trace des w(abc)}
          \end{eqnarray} 
which gives back (\ref{trace de uu}) for $a = a'=0$.
 
The product of the operators $w_{a b c}$ and $w_{a' b' c'}$ reads 
     \begin{eqnarray}
w_{a b c} w_{a' b' c'} = w_{a'' b'' c''} \qquad 
a'' = a \oplus a' \ominus cb' \qquad 
b'' = b \oplus b' \qquad 
c'' = c \oplus c'.
     \label{produit des w(abc)}
     \end{eqnarray}   
The set $\{ w_{a b c}: a, b, c \in \mathbb{Z}_d \}$ can be endowed with a group structure. In the detail, 
we have the following 

\bigskip

\noindent {\bf Result 14.} {\it The set $\{ w_{abc} : a, b, c \in \mathbb{Z}_d \}$, endowed with the internal 
law (\ref{produit des w(abc)}), is a finite group of order $d^3$. This nonabelian group, 
noted $\Pi_d$ and called generalized Pauli group in $d$ dimensions, is nilpotent (hence 
solvable) with a nilpotency class equal to 2. The group $\Pi_d$ is isomorphic to a 
subgroup of $U(d)$ for $d$ even or $SU(d)$ for $d$ odd. It has $d(d+1)-1$ conjugacy classes ($d$ classes containing each 
1 element and $d^2 - 1$ classes containing each $d$ elements) and $d(d+1) - 1$ classes of irreducible 
representations ($d^2$ classes of dimension 1 and $d - 1$ classes of dimension $d$).} 

\bigskip

A faithful three-dimensional representation of $\Pi_d$ is provided with the application 
         \begin{eqnarray}
\Pi_d \to GL(3 , \mathbb{Z}_d) : w_{abc} \mapsto 
\pmatrix{
1   &    0  &      0   \cr
b   &    1  &      0   \cr
a   &   -c  &      1   \cr
}.
          \label{faithful rep} 
          \end{eqnarray} 
This is reminiscent of the Heisenberg-Weyl group \cite{Wolf, Wolf2, Wolf3, Wolf4, Terras, Howe}. Indeed, 
the group  $\Pi_d$  can be considered as a discretization $HW(\mathbb{Z}_d)$ of the 
Heisenberg-Weyl group $HW(\mathbb{R})$, a three-parameter Lie group. The 
Heisenberg-Weyl group $HW(\mathbb{R})$, also called the Heisenberg group or Weyl 
group, is at the root of quantum mechanics. It also plays an important role in 
symplectic geometry. The group $\Pi_d$ was discussed 
by \v{S}\v{t}ov\'{\i}\v{c}ek and Tolar \cite{Tolar1} in connection with quantum 
mechanics in a discrete space-time, by Balian and Itzykson \cite{Balian} in connection with 
finite quantum mechanics, by Patera and Zassenhaus  
\cite{PateraZassenhaus} in connection with gradings of simple Lie algebras of 
type $A_{n-1}$, and by Kibler \cite{KibJPhysA} in connection with Weyl pairs and 
the Heisenberg-Weyl group. Note 
that the discrete version $HW(\mathbb{Z})$ of $HW(\mathbb{R})$ was used for 
an analysis of the solutions of the Markoff equation \cite{Perrine}. Recently, 
the discrete version $HW[\mathbb{Z}_p \times (\mathbb{Q}_p / \mathbb{Z}_p) ]$ 
was introduced for describing ($p$-adic) quantum systems with positions in 
$\mathbb{Z}_p$ and momenta in $\mathbb{Q}_p / \mathbb{Z}_p$ \cite{Vourdas2008}. 

    As far as $HW(\mathbb{Z}_d)$ is concerned, it is to be observed that a Lie 
algebra $\pi_d$ can be associated with the finite group $\Pi_d$. This can be 
seen by considering the Frobenius algebra of $\Pi_d$ (see \cite{Gamba} for the 
definition of the Lie algebra associated with an arbitray finite group). Then, 
the Lie brackets of $\pi_d$ are 
     \begin{eqnarray}
[ w_{a b c} , w_{a' b' c'} ]_- = w_{\alpha \beta \gamma} - w_{\alpha' \beta' \gamma'} 
     \label{commutateur des w(abc)}
     \end{eqnarray}
with 
     \begin{eqnarray}
\alpha  = a \oplus a' \ominus cb' \ \ \
\beta   = b \oplus b' \ \ \
\gamma  = c \oplus c' \ \ \
\alpha' = \alpha \oplus cb' \ominus bc' \ \ \
\beta'  = \beta \ \ \
\gamma' = \gamma.
     \label{parameters of commutateur des w(abc)}
     \end{eqnarray}       
The algebra $\pi_d$, of dimension $d^3$, is not semisimple. It can be decomposed 
as the direct sum
\begin{eqnarray}
\pi_d \simeq \biguplus_{1}^{d^2} {u}(1) \biguplus_{1}^{d-1} {u}(d)  
\label{decomposition of pd} 
\end{eqnarray}
which contains $d^2$ Lie algebras isomorphic to $u(1)$ and $d-1$ 
Lie algebras isomorphic to $u(d)$. The Lie algebra $u(d)$ spanned by the set 
$\{ u_{ab} : a,b \in \mathbb{Z}_d \}$ is one of the subalgebras of $\pi_d$. 

The group $\Pi_d$ (noted $P_d$ in \cite{KibJPhysA}) should not be confused with 
the Pauli group ${\cal P}_n$ on $n$ qubits spanned by $n$-fold tensor products 
of $i \sigma_0 \equiv i I_2$, $\sigma_x$ and $\sigma_z$ used in quantum 
information and quantum computation \cite{Nielsen, Jozsa}. The Pauli group 
${\cal P}_n$ has $4^{n+1}$ elements. It is used as an error group in quantum 
computing. The normaliser of ${\cal P}_n$ in $SU(2^n)$, known 
as the Clifford (or Jacobi) group $Cli_n$ on $n$ qubits, a group of 
order $2^{n^2 + 2n +3} \prod_{j = 1}^n (4^j - 1)$, is of great interest in the context 
of quantum corrector codes 
\cite{Grassl, Gottesman, Calderbank, Clark, Appleby, Flammia, Cormick}. In addition to $Cli_n$, proper 
sugroups of $Cli_n$ having ${\cal P}_n$ as an invariant subgroup are relevant for 
displaying quantum coherence \cite{PlaJor}. The distinction bewteen ${\cal P}_n$ and $\Pi_d$ 
can be clarified by the example below which shows that $\Pi_2$ is not isomorphic 
to ${\cal P}_1$. In a parallel way, it can be proved that the 
groups $\Pi_4$ and ${\cal P}_2$ (both of order 64) are distinct. 

\bigskip

\noindent {\bf Example 9:} $d=2$.  The simplest example of $\Pi_d$ occurs for $d=2$. The group $\Pi_2$ has 
8 elements ($\pm I$, $\pm x$, $\pm y := \pm xz$, $\pm z$) and is isomorphic to the dihedral group $D_4$. It 
can be partitioned into 5 conjugation classes 
($\{ I \}, \{ -I \}, \{ x , -x \}, \{ y , -y \}, \{ z , -z \}$) and possesses 
5 inequivalent irreducible representations (of dimensions 1, 1, 1, 1 and 2). 
The two-dimensional irreducible representation corresponds to 
\begin{eqnarray}
\pm I \mapsto \pm   \sigma_0 \qquad 
\pm x \mapsto \pm   \sigma_x \qquad 
\pm y \mapsto \mp i \sigma_y \qquad 
\pm z \mapsto \pm   \sigma_z 
\end{eqnarray}
in terms of the Pauli matrices $\sigma_{\lambda}$ with $\lambda = 0, x, y, z$. 
The elements $e_1 := x$, $e_2 := y$ and $e_3 := z$ of $\Pi_2$ span the four-dimensional algebra 
$A(1, -1, 0) \equiv \mathbb{N}_1$, the algebra of \emph{hyperbolic quaternions} 
(with $e_1^2 = - e_2^2 = e_3^2 = 1$ instead of $e_1^2 = e_2^2 = e_3^2 = -1$ as for usual quaternions).
This associative and noncommutative algebra is a singular division algebra. The 
algebra $\mathbb{N}_1$ turns out to be a particular Cayley-Dickson algebra 
$A(c_1, c_2, c_3)$ \cite{Lambert}. Going back to $\Pi_2$, we 
see that not all the subgroups of $\Pi_2$ are invariant. The 
group $\Pi_2$ is isomorphic to the group of \emph{hyperbolic quaternions} 
rather than to the group $Q$ of \emph{ordinary quaternions} for which all 
subgroups are invariant (the group $Q$ can be realized with the help of the 
matrices $\pm \sigma_0$, $\pm i \sigma_x$, $\pm i \sigma_y$, $\pm i \sigma_z$). Like 
$Q$, the group $\Pi_2$ is ambivalent and 
simply reducible in the terminology of Wigner \cite{WignerSR}. Indeed, 
$\Pi_2$ is the sole generalized Pauli group that is ambivalent. 

To end up with this example, let us examine the connection between $\Pi_2$ and 
the 1 qubit Pauli group ${\cal P}_1$. The group ${\cal P}_1$ has 16 elements 
($\pm \sigma_{\lambda}, \pm i \sigma_{\lambda}$ with $\lambda = 0, x, y, z$). Obviously, 
$\Pi_2$ is a subgroup of index 2 (necessarily invariant) of ${\cal P}_1$. The group  
${\cal P}_1$ can be considered as a double group of $\Pi_2$ or $Q$ in the sense that ${\cal P}_1$ 
coincides with $\Pi_2 \bigcup i \Pi_2 \equiv Q \bigcup i Q$ in terms of sets. Therefore, the group table of 
${\cal P}_1$ easily follows from the one of $\Pi_2$ or $Q$. As a result, the numbers of conjugation 
classes and irreducible representation classes are doubled when passing from $\Pi_2$ or $Q$
to ${\cal P}_1$. 

\section{Closing remarks}

Starting from a nonstandard approach to angular momentum and its transcription in terms of 
representation of $SU(2)$, we derived (in an original and unified way) some results about 
unitary operator bases and their connection to unitary groups, Pauli groups and quadratic 
sums. These results (either known or formulated in a new way) shed some light on the 
importance of the polar decomposition of $su(2)$ and cyclic groups for the study of unitary 
operator bases and their relationship with unitary groups and Pauli groups. In particular, 
the latter decomposition and the quadratic discrete Fourier transform introduced in this work 
make it possible to generate in prime dimension a complete set of MUBs given by a single formula 
(Eq.~(\ref{vecteur a alpha})). From the point of view of the representation theory, it would be 
interesting to find realisation on the state vectors (\ref{vecteur a alpha}) on the sphere 
$S^2$ thus establishing a contact between the $\{ j^2 , v_{0a} \}$ scheme and special functions.  

To close this paper, let us be mention two works dealing with {\it \`a la} Schwinger unitary 
operator bases in an angular momentum scheme. In Ref.~\cite{Marchiolli}, unitary operator 
bases and standard (discrete) quantum Fourier transforms in an angular momentum framework 
proved to be useful for spin tunneling. In addition, $d$-dimensional generalized Pauli 
matrices applied to modified Bessel functions were considered in an angular momentum 
approach with $j = (d-1)/2 \to \infty$ \cite{Fujii}.
     
     \subsection*{Appendix A: A polar decomposition of $su(2)$}

In addition to the operator $v_{ra}$, a second linear operator 
is necessary to define a polar decomposition of 
$SU(2)$. Let us introduce the Hermitian operator $h$ through 
        \begin{eqnarray}
   h := \sum_{m = -j}^j {\sqrt{ (j+m)(j-m+1) }} |j , m \rangle \langle j , m |. 
        \label{definition of h}
        \end{eqnarray}
Then, it is a simple matter of calculation to show that the three operators 
	  \begin{eqnarray}
  j_+ := h           v_{ra}    \qquad  
  j_- := (v_{ra})^{\dagger} h  \qquad 
  j_z := \frac{1}{2} \left[ h^2 - (v_{ra})^{\dagger} h^2 v_{ra} \right]
          \end{eqnarray}
satisfy the ladder equations
     \begin{eqnarray}
  j_+ |j , m \rangle   & = & q^{+(j - m + s - 1/2)a}
  {\sqrt{ (j - m)(j + m+1) }} 
  |j , m + 1 \rangle 
  \\
  j_- |j , m \rangle   & = & q^{-(j - m + s + 1/2)a}
  {\sqrt{ (j + m)(j - m+1) }} 
  |j , m - 1 \rangle 
     \end{eqnarray} 
and the eigenvalue equation 
               \begin{eqnarray} 
  j_z   |j , m \rangle = m |j,m \rangle
               \end{eqnarray}
where $s = 1/2$. Therefore, the operators $j_+$, $j_-$ and $j_z$ satisfy the commutation 
relations
     \begin{eqnarray}
  \left[ j_z,j_{+} \right]_- = + j_{+}  \qquad 
  \left[ j_z,j_{-} \right]_- = - j_{-}  \qquad 
  \left[ j_+,j_- \right]_- = 2j_z 
     \label{adL su2}
     \end{eqnarray}
and thus span the Lie algebra of $SU(2)$.

The latter result does not depend on the parameters $r$ and $a$. However, the action 
of $j_+$ and $j_-$ on $|j , m \rangle$ on the space ${\cal E} (2j+1)$ depends on $a$;  
the usual Condon and Shortley phase convention used in spectroscopy
corresponds to $a = 0$. The writing of the ladder operators $j_+$ and $j_-$ in
terms of $h$ and $v_{ra}$ constitutes a two-parameter polar decomposition of 
the Lie algebra of $SU(2)$. This decomposition is an alternative to the polar 
decompositions obtained independently in \cite{Vourdas04, Leblond, Vourdas90, Chaichian, Ellinas}. 

\subsection{Appendix B: A quon approach to $su(2)$}

Following \cite{DaoHasKib}, we define a quon algebra or $q$-deformed oscillator 
algebra for $q$ a root of unity. The three operators $a_-$, $a_+$ and $N_a$ such that
              \begin{eqnarray}
  a_- a_+  - q a_+ a_- = I \qquad
  [ N_a , a_{\pm} ]_- = {\pm} a_{\pm} \qquad
  (a_{\pm})^k = 0 \qquad 
  N_a ^{\dagger} = N_a
              \label{Aq(x)}
              \end{eqnarray}
where
              \begin{eqnarray}
q := \exp \left( \frac{2\pi { i }}{k} \right) \qquad 
k \in \mathbb{N} \setminus \{ 0,1 \}
              \label{def(q,k)}
              \end{eqnarray}
define a quon algebra or $q$-deformed oscillator algebra denoted as 
$A_q(a_-, a_+, N_a)$. The operators $a_-$ and $a_+$ are referred to 
as quon operators. The operators $a_-$, $a_+$ and $N_a$ are called 
annihilation, creation and number operators, respectively.

Let us consider two commuting quon algebras $A_q(a_-, a_+, N_a) \equiv A_q(a)$ 
with $a = x, y$ corresponding to the same value of the deformation parameter 
$q$. Their 
generators satisfy equations (\ref{Aq(x)}) and 
(\ref{def(q,k)}) with $a = x, y$ and $[X, Y]_- = 0$ 
for any $X$ in $A_q(x)$ and 
    any $Y$ in $A_q(y)$. Then, let us look for Hilbertian representations 
of $A_q(x)$ and $A_q(y)$ on the $k$-dimensional Hilbert spaces 
${\cal F}(x)$ and 
${\cal F}(y)$ spanned by the orthonormal bases   
$\{ | n_1 ) : n_1 = 0, 1, \ldots, k-1 \}$ and 
$\{ | n_2 ) : n_2 = 0, 1, \ldots, k-1 \}$, respectively. We easily
obtain the representations defined by 
      \begin{eqnarray}
  && x_+ |n_1) = |n_1 + 1)                      \qquad  
     x_+ |k-1) = 0                              \nonumber \\ 
  && x_- |n_1) = \left[ n_1   \right]_q |n_1-1) \qquad   
     x_- |0)   = 0                              \nonumber \\
  && N_x |n_1) = n_1 |n_1)
      \end{eqnarray}
and
      \begin{eqnarray}
  && y_+ |n_2) = \left[ n_2+1 \right]_q |n_2+1) \qquad  
     y_+ |k-1) = 0                              \nonumber \\ 
  && y_- |n_2) = |n_2 - 1)                      \qquad  
     y_- |0) = 0                                \nonumber \\  
  && N_y |n_2) = n_2 |n_2)                      
      \end{eqnarray}
for $A_q(x)$ and $A_q(y)$, respectively. 

The cornerstone of this approach is to define the two linear operators
              \begin{eqnarray}  
  h := {\sqrt {N_x \left( N_y + 1 \right) }} \qquad v_{ra} := s_x s_y 
  \label{h and v_a n1}
              \end{eqnarray}
with  
\begin{eqnarray}
   s_x &:=& q^{ a (N_x + N_y) / 2 } x_{+} + 
   {e} ^{ {i} \phi_r / 2 }  {1 \over 
  \left[ k-1 \right]_q!} (x_{-})^{k-1} 
  \\
   s_y &:=& y_{-}  q^{- a (N_x - N_y) / 2 }+ 
   {e} ^{ {i} \phi_r / 2 }  {1 \over 
  \left[ k-1 \right]_q!} (y_{+})^{k-1} 
\end{eqnarray}
where 
          \begin{eqnarray}
 a \in \mathbb{R} \qquad \phi_r = \pi (k-1) r \qquad r \in \mathbb{R}. 
          \end{eqnarray}
The operators $h$ and $v_{ra}$ act on the states 
\begin{eqnarray}
| n_1 , n_2 ) := | n_1) \otimes | n_2 ) 
\end{eqnarray} 
of the $k^2$-dimensional space ${\cal F}_k := {\cal F}(x) \otimes {\cal F}(y)$. It 
is immediate to show that the action of $h$ and $v_{ra}$ on ${\cal F}_k$ is given by
          \begin{eqnarray}
  h |n_1 , n_2) = {\sqrt{ n_1 (n_2 + 1) } |n_1 , n_2)} 
  \qquad n_i = 0, 1, 2, \cdots, k-1 
  \qquad i=1,2 
  \label{action de h sur n1n2}
          \end{eqnarray}
and
          \begin{eqnarray}
  v_{ra} |n_1 , n_2) = q^{a n_2} |n_1+1 , n_2-1)
  \qquad n_1 \not = k-1 
  \qquad n_2 \not = 0 
  \label{action1 de v_a sur n1n2}
          \end{eqnarray}
          \begin{eqnarray}
  v_{ra} |k-1 , n_2) =  {e} ^{ {i} {\phi}_r / 2 } 
  q^{- a (k - 1 - n_2) / 2}
  |0 , n_2 - 1) 
  \qquad n_2 \not= 0 
          \end{eqnarray}
          \begin{eqnarray}
  v_{ra} |n_1 , 0) =  {e} ^{ {i} {\phi}_r / 2 } 
  q^{ a (k + n_1) / 2 }
  |n_1 + 1 , k-1) 
  \qquad n_1 \not= k-1 
          \end{eqnarray}
          \begin{eqnarray}
  v_{ra} |k-1 , 0) =  {e} ^{{i} {\phi}_r} 
  |0 , k-1). 
  \label{action2 de v_a sur n1n2}
          \end{eqnarray}
The operators $h$ and $v_{ra}$ satisfy interesting properties: the 
operator $h$ is Hermitian and the operator $v_{ra}$ is unitary.

We now adapt the trick used by Schwinger \cite{Schwingertrick} in his approach 
to angular momentum via a coupled pair of harmonic oscillators. This can be 
done by introducing two new quantum numbers $J$ and $M$ defined by
          \begin{eqnarray}
  J := {1 \over 2} \left( n_1+n_2 \right) \quad  
  M := {1 \over 2} \left( n_1-n_2 \right) \Rightarrow
  |J M \rangle := |J + M , J-M) = |n_1 , n_2).
          \end{eqnarray}  
Note that 
          \begin{eqnarray}
j := \frac{1}{2}(k-1)
          \end{eqnarray}  
is an admissible value for $J$. Then, let us consider the $k$-dimensional
subspace $\epsilon(j)$ of the $k^2$-dimensional space 
${\cal F}(x) \otimes {\cal F}(y)$ spanned by the basis 
$\{ |j , m \rangle : m = j, j-1, \ldots, -j \}$. 
We guess that $\epsilon(j)$ is a space of constant angular momentum $j$. As a 
matter of fact, we can check that $\epsilon(j)$ is stable under $h$ and 
$v_{ra}$. In fact, the action of the operators $h$ and $v_{ra}$ on the 
    subspace $\varepsilon (j)$ of ${\cal F}_k$ can be described by 
          \begin{eqnarray}
  h |j , m \rangle = {\sqrt{ (j+m)(j-m+1) }} |j , m \rangle 
          \end{eqnarray} 
and
          \begin{eqnarray}
  v_{ra} |j , m \rangle = \delta_{m,j} {e}^{{i} 2 \pi j r} |j , -j \rangle + 
  \left( 1 - \delta_{m,j} \right) q^{(j-m)a} |j , m+1 \rangle 
          \end{eqnarray}
in agreement with Eq.~(\ref{definition of h}) and with the master equation 
(\ref{definition of vra}).

\baselineskip = 0.50 true cm

\end{document}